\documentclass[twocolumn]{aastex631}
\usepackage{graphics,epsf}
\usepackage{amsmath}                
\usepackage{amsfonts}               
\usepackage{amssymb}                
\usepackage{epsfig}                 
\usepackage{appendix}
\usepackage{graphicx}
\usepackage{float}
\usepackage{color}
\usepackage{multirow}
\usepackage{colortbl}
\usepackage[para,online,flushleft]{threeparttable}

\hypersetup{citecolor=blue, 
            linkcolor=red, 
            menucolor=blue, 
            urlcolor=blue}  

\newcommand{\cm}{{~\rm cm}}
\newcommand{\m}{{~\rm m}}
\newcommand{\km}{{~\rm km}}
\newcommand{\s}{{~\rm s}}
\newcommand{\h}{{~\rm h}}

\newcommand{\g}{{~\rm g}}
\newcommand{\G}{{~\rm G}}
\newcommand{\K}{{~\rm K}}
\newcommand{\erg}{{~\rm erg}}
\newcommand{\yr}{{~\rm yr}}

\begin{document}

\title{Jet-powered turbulence in common envelope evolution}

\author{Shlomi Hillel}
\affiliation{Department of Physics, Technion, Haifa, 3200003, Israel; \\ 	
shlomi.hillel@gmail.com, ronsr@physics.technion.ac.il, soker@physics.technion.ac.il}

\author{Ron Schreier}
\affiliation{Department of Physics, Technion, Haifa, 3200003, Israel; \\ 	
shlomi.hillel@gmail.com, ronsr@physics.technion.ac.il, soker@physics.technion.ac.il}

\author[0000-0003-0375-8987]{Noam Soker}
\affiliation{Department of Physics, Technion, Haifa, 3200003, Israel; \\ 	
shlomi.hillel@gmail.com, ronsr@physics.technion.ac.il, soker@physics.technion.ac.il}

\begin{abstract}
We conduct a three-dimensional hydrodynamical simulation of a common envelope evolution (CEE) where a neutron star (NS) spirals-in inside the envelope of a red supergiant (RSG) star in a predetermined orbit. We find that the jets shed pairs of vortices in an expanding spiral pattern, inflate two expanding spirally-shaped low-density bubbles, one above and one below the equatorial plane, and deposit angular momentum to the envelope. In the simulation we do not include the gravity of the NS such that all effects we find are solely due to the jets that the spiralling-in NS launches. The angular momentum that the jets deposit to the envelope is of the same order of magnitude as the orbital angular momentum and has the same direction. The turbulence that the jets induce in the common envelope might play a role in transporting energy and angular momentum. The jet-deposited energy that is radiated away (a process not studied here) leads to a transient event that is termed common envelope jets supernova (CEJSN) and might mimic an energetic core collapse supernova.  The turbulence and the spiral pattern that we explore here might lead to bumps in the late light curve of the CEJSN when different segments of the ejected envelope collide with each other. This study emphasises the roles that jets can play in CEE (including jets launched by black hole companions) and adds to the rich variety of processes in CEJSN events.
\end{abstract}


\section{Introduction} 
\label{sec:intro}

Compact objects that spiral-in inside the extended envelope of giant stars, i.e., common envelope evolution (CEE), might accrete mass through an accretion disk and launch jets. This is very likely to be the case when the compact objects are neutron stars (NSs) (e.g., \citealt{ArmitageLivio2000, Chevalier2012, LopezCamaraetal2019, LopezCamaraetal2020MN}) and black holes (BHs), and to some degree also main sequence stars, as some planetary nebulae hint at (e.g., \citealt{BlackmanLucchini2014}; for a review see \citealt{Soker2016Rev}) and theory supports (e.g., \citealt{Soker2023MS}). A key aspect is that the jets regulate the accretion rate onto the compact object and by that the energising process of the CEE. Namely, the jets operate in a negative feedback mechanism, e.g., \cite{Soker2016Rev} for a review, \cite{Gricheneretal2021} for one-dimensional (1D) simulations, and \cite{Hilleletal2022FB} for 3D simulations. 
There is also a positive feedback component where the jets remove energy from the accreting body vicinity, thereby facilitating accretion at a higher rate (e.g., \citealt{Shiberetal2016, Chamandyetal2018, LopezCamaraetal2019, LopezCamaraetal2020MN}). 

Most 3D simulations of the CEE do not include jets (e.g., \citealt{Passyetal2012, RickerTaam2012, Nandezetal2014, Staffetal2016MN, Kuruwitaetal2016, Ohlmannetal2016a,  Iaconietal2017b, Chamandyetal2019, LawSmithetal2020, GlanzPerets2021a, GlanzPerets2021b, GonzalezBolivar2022, Lauetal2022a, Lauetal2022b, Chamandyetal2023} for a very limited list; for a recent thorough review with many more references see \citealt{RoepkeDeMarco2023}). The limited number of 3D hydrodynamical simulations of the CEE (and the grazing envelope evolution) that do include jets launched by the companion (e.g., \citealt{MorenoMendezetal2017, ShiberSoker2018, LopezCamaraetal2019, Schreieretal2019inclined, Shiberetal2019, LopezCamaraetal2020MN, LopezCamaraetal2022, Zouetal2022, Schreieretal2023}) are far from revealing all aspects of jet-powered CEE (see \citealt{Soker2022Rev} for a review of processes due to jets that NS/BH launch in CEE and possible outcomes). A different class of simulations (e.g., \citealt{Zouetal2020, Morenoetal2022, Ondratscheketal2022}) study collimated outflows from the distorted envelope at the final phases of the CEE (similar to the suggestion by \citealt{Soker1992}), but this setting is not related to the present study.   

In this study we extend our exploration of CEE with jets that a NS companion launches as it orbits inside the envelope of a red supergiant (RSG) star. Using the hydrodynamical code {\sc flash} (section \ref{sec:Numerical}) we simulate the effect of the jets that a NS launches as it spirals-in. Before we present the effects of the jets in sections \ref{sec:Turbulence} and \ref{sec:AngularMomentum} we discuss the construction of the 3D stellar model (section \ref{sec:StellarModel}). We summarise our main results in section \ref{sec:Summary}.

\section{The numerical setup}
\label{sec:Numerical}
Our numerical procedure is similar to that in our previous paper \citep{Schreieretal2023}, but in some simulations we employ higher resolution and, most importantly, we set the NS to spiral-in into the RSG envelope rather than have a constant orbit. We therefore do not describe all numerical details here, but rather only the essential ingredients. 

\subsection{The stellar model and the NS orbit}
\label{subsec:StellarOrbit}

Using the \texttt{MESA} one-dimensional (1D) stellar evolution code \citep{Paxtonetal2011, Paxtonetal2013, Paxtonetal2015, Paxtonetal2018, Paxtonetal2019} we evolve a zero-age-main-sequence star of metalicity $Z=0.02$ and mass $M_{\rm 1,ZAMS}=15 M_\odot$ to the RSG phase. 
We place the RSG stellar model at the centre of the 3D hydrodynamical numerical grid at an age of $1.1\times10^6 \yr$ 
when its radius is $R_{\rm RSG}=881\,R_{\odot}$, its mass is $M_1=  12.5 M_\odot$, and its effective temperature is $T_{\rm eff}= 3160K$. We use the 3D  hydrodynamical code {\sc flash} \citep{Fryxelletal2000} with fully ionised pure hydrogen.   
We set the numerical grid cells outside the stellar model to have a very low density $\rho_{\rm grid,0} = 2.1 \times 10^{-13} \g \cm^{-3}$ and have a temperature of $T_{\rm grid,0}= 1100 \K$.  

To save numerical time we do not calculate the flow in the inner $20\%$ of the stellar radius, $R_{\rm in} = 176\,R_{\odot}$. We rather take this inert core to be a central sphere with constant density, pressure and temperature. We fully take into account the gravity of the inert core in the entire grid. We also include the gravity of the envelope as it is at $t=0$. Namely, the gravity of the stellar model is constant throughout the evolution and at each radius equals that of the stellar model at $t=0$. We study the behaviour of this model in the 3D grid in section \ref{sec:StellarModel}. 

We assume a common envelope jet supernova (CEJSN) event where the NS spirals-in inside the RSG accretes mass and launches jets. We do not include the gravity of the NS nor the orbital energy or angular momentum. We preset the orbit of the NS as follows. The NS spiral-in from $a_{\rm i} = 850\,R_{\odot}$
to $a_{\rm SR} = 300\,R_{\odot}$ in a time period of 3 years that mimics the plunge-in phase of the CEE, after which it stays in a constant circular orbit which mimics the self-regulated CEE phase (for that the subscript `SR'). The radial velocity during the plunge-in phase is constant, whereas the orbital velocity is Keplerian. 

\subsection{The numerical grid}
\label{subsec:Grid}

Our simulations are performed on a cubic Cartesian computational grid with a side of $L_{\rm G} = 5 \times 10^{14} \cm$.  We set outflow conditions on all boundary surfaces of the 3D grid. 
Adaptive mesh refinement (AMR) is employed with a refinement criterion of a modified L\"ohner error estimator (with default parameters) on the $z$-component of the velocity.
The gas in the whole computational domain is an ideal gas with an adiabatic index of $\gamma=5/3$ including radiation pressure.
The centre of the RSG is fixed at the origin. The smallest cell size in the simulation is  $L_{\rm G}/128 = 3.90625 \times 10^{12} \cm$.

In one simulation of the 3D stellar model without jets (section \ref{sec:StellarModel}) we use  higher resolution where all grid cells have
the same size of $L_{\rm G}/512 = 9.77 \times 10^{11} \cm$.

\subsection{Jet-launching procedure}
\label{subsec:Jets}

Our limited computational resources force us to employ a sub-grid procedure to study the effects of the jets. In the new scheme that we developed in \cite{Schreieretal2023} we inject energy and momentum that serve as the jets' parameters rather than the jets' energy and velocity. This sub-grid procedure does not change the mass of any computational grid cell. We do not add or remove mass in the launching procedure. We only change the velocity and internal energy of the already existing mass in the volume where we inject the jets' energy. 

The power of the two jets should vary with density $\rho(a)$ at the location of the NS $a$ and the relative velocity of the NS inside the RSG envelope, $v_{\rm rel}(a)$,  according to   
\begin{equation}
\dot E_{\rm 2j} = \zeta \frac{G M_{\rm NS}}{R_{\rm NS}}\dot M_{\rm BHL,0},
\label{eq:JetsPower}    
\end{equation}
where  
\begin{equation}
\dot M_{\rm BHL,0} = \pi \rho(a) v_{\rm rel} (a)
\left[ \frac{2 G M_{\rm NS}}{v^2_{\rm rel}(a)} \right]^2
\label{eq:MBHL}
\end{equation}
is the Bondi-Hoyle-Lyttleton (BHL) mass accretion rate from the unperturbed envelope and $\zeta \simeq 0.002 - 0.005$ (\citealt{Gricheneretal2021, Hilleletal2022FB}). 
For the mass and radius of the NS we take $M_{\rm NS}=1.4 M_\odot$ and $R_{\rm NS}=12 \km$, respectively.  
In the mass accretion rate expression we neglect the sound speed in the envelope, which reduces the accretion rate, and the envelope rotation, which  reduces the relative velocity and therefore increases the accretion rate. 
Because we do not include the gravity of the NS, 
the properties of the NS enter through equations (\ref{eq:JetsPower}) and (\ref{eq:MBHL}). 
Because of numerical limitations we mostly simulate jets with powers smaller than what equation (\ref{eq:JetsPower}) gives. 

We set the NS to spiral-in from $a = 850\,R_{\odot}$
to $a = 300\,R_{\odot}$ in a time span of 3 years.
The local unperturbed density increases from
$\rho (850\,R_{\odot}) = 3.3\times 10^{-9} \g \cm^{-3}$
to $\rho (300\,R_{\odot}) = 5.9\times 10^{-8} \g \cm^{-3}$. The power of the two jets changes accordingly from
$\dot E_{2j} = 3.3\times 10^{40} \erg \s^{-1}$ to
$3\times 10^{41} \erg \s^{-1}$,
keeping $\zeta =2\times 10^{-5}$ constant.

As stated before, we cannot resolve the launching region of the jets. Instead, we insert in the grid the two opposite jet-envelope interaction zones near the NS. 
We take these zones to be two cylinders touching each other at the orbital plane, i.e., they form one cylinder with the NS at its centre. The jets' axis is the axis of the cylinder. The base of the cylinder has a radius of $4 \times 10^{12} \cm$ and the total height is $14 \times 10^{12} \cm$, i.e., $7 \times 10^{12} \cm$ on each side of the equatorial plane.
The momentum that we deposit inside the cylinder is in the direction of the axis of the cylinder and away from the NS, i.e., two zones perpendicular to the equatorial plane with two opposite outflows away from the equatorial plane. 

The total momentum discharge rate is $\dot{P}_{\rm 2j} \equiv \vert \dot{P}_{\rm 1} \vert + \vert \dot{P}_{\rm 2} \vert$, where the indices stand for the two opposite jets.
The value of $\dot{P}_{\rm 2j}$ is related to the power of the two jets by postulating that the jets are launched at a constant speed of $v_{\rm j}=5 \times 10^4 \km \s^{-1}$, i.e.,
\begin{equation}
\dot{P}_{\rm 2j} = \frac{2 \dot E_{\rm 2j}}{v_{\rm j}}.
\label{eq:JetsMomentum}
\end{equation}

In each time step $\Delta t$ we first change the velocity inside the jet-injection cylinder as a result of the momentum that we add. To each grid cell inside the jet-injection cylinder we add a momentum of 
\begin{equation}
\Delta p_{\rm c}= f_{\rm V,c} \dot P_{\rm 2j} \Delta t,
\label{eq:Pnewcell}
\end{equation} 
where $f_{\rm V,c}$ is the fraction of the cylinder volume that the cell occupies. Using the mass in each cell and its previous velocity we compute the new velocity of the cell. 
In the second step we compute the new energy of each grid cell inside the jet-injection cylinder 
\begin{equation}
E_{\rm new,c}=E_{\rm old,c}+f_{\rm V,c} \dot E_{\rm 2j} \Delta t, 
\label{eq:Enewcell}
\end{equation}
where $E_{\rm old,c}$ is the old total energy in the cell, including only the kinetic energy and the internal energy because the gravitational energy does not change during the jet-injection process. Because we know already the new kinetic energy in the cell, $E_{\rm new,kin,c}$, from the new velocity as we calculate from the new momentum (equation \ref{eq:Pnewcell}), equation (\ref{eq:Enewcell}) serves to calculate the new thermal energy in each cell in the jet-injection cylinder. Namely, the third step is taking $E_{\rm new,therm,c}=E_{\rm new,c}-E_{\rm new,kin,c}$ in each cell inside the jet-injection cylinder. 

This scheme conserves mass, momentum, and energy (see \citealt{Schreieretal2023} for more numerical details).

\section{The three-dimensional stellar model} 
\label{sec:StellarModel}

Our goal in this section is to reveal the behaviour of the 3D model that we transported from the 1D model of \textsc{mesa} (section \ref{subsec:StellarOrbit}) and consider the implications for the building of 3D giant models. For that goal we follow the evolution of the 3D stellar model without jets. We recall that in the 3D model we use there is a numerical spherical inert core of radius $R_{\rm in} = 176\,R_{\odot}$ (section \ref{subsec:StellarOrbit}) that saves us expensive computational time. Its gravity is fully included in the simulations. We set $t=0$ when we start the 3D simulations, at which time the RSG radius is $R_{\rm RSG}=881R_{\odot}$ and the RSG mass is $M_1=12.5M_\odot$. 

We present the results of the regular resolution that we use in the simulations with jets (section \ref{sec:Turbulence}) and of a simulation with  higher resolution. The smallest cell size in the regular resolution is  $L_{\rm G}/128 = 3.90625 \times 10^{12} \cm$, while in the high resolution simulation all  cells have the same size of $L_{\rm G}/512 = 9.77 \times 10^{11} \cm$.  
(We have no computer resources to simulate jets with the high-resolution grid; the simulations without jets are much faster than those with jets, and we can afford the high-resolution grid.) 

There are two timescales that we will refer to in the discussion to follow, 
\begin{equation}
P_{\rm Kep}=2.35  \yr, \quad {\rm and} \quad  P_{\rm D}\equiv (G \bar \rho)^{-1/2} =0.76  \yr,
\label{eq:TimeScales}
\end{equation}
where $P_{\rm Kep}$ is the Keplerian orbital time on the surface of the initial RSG model, $P_{\rm D}$ is the dynamical time,  and  
$\bar \rho$ is the average density of the initial RSG model. 

\begin{figure*} 
\centering
\includegraphics[width=0.45\textwidth]{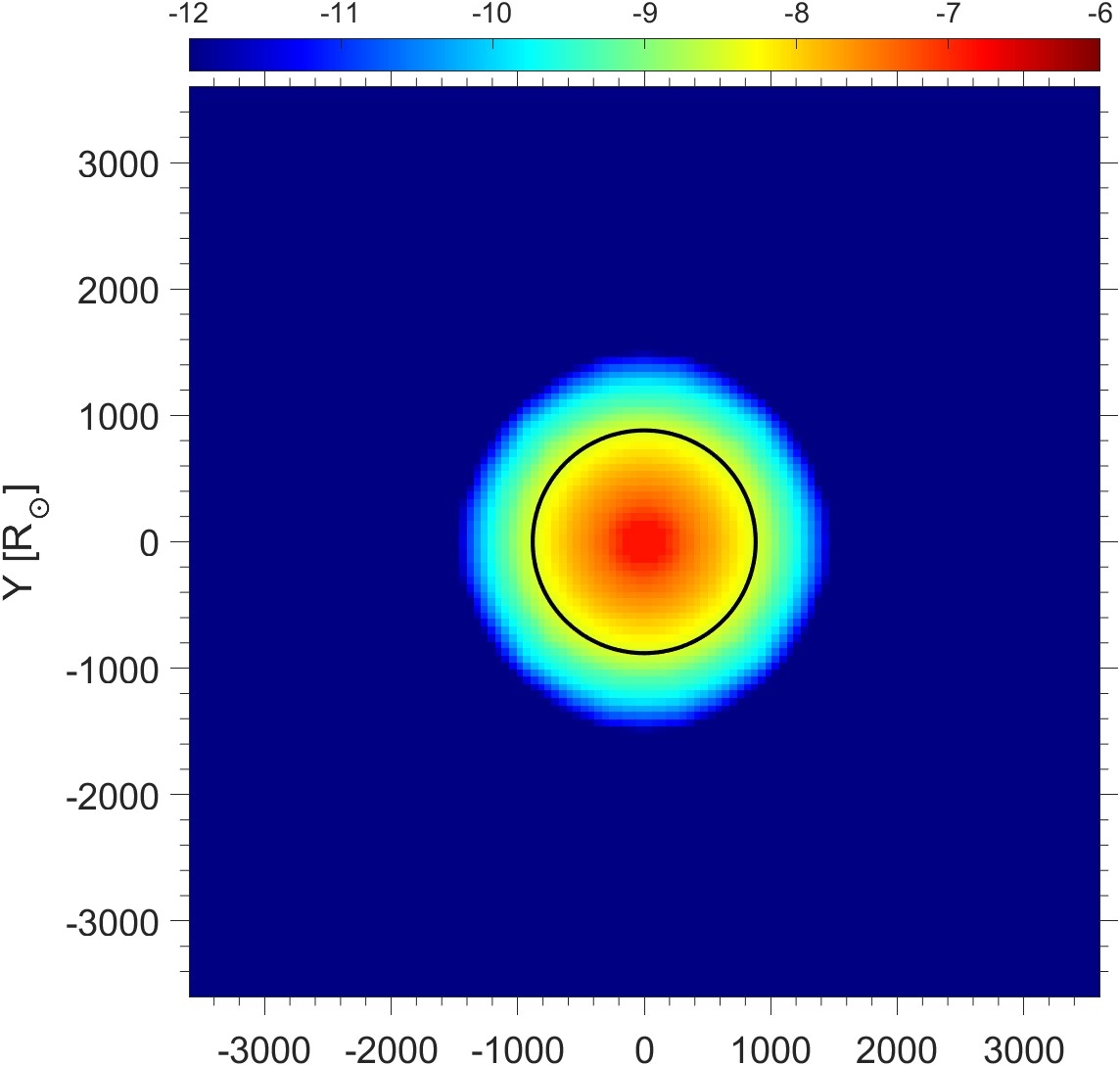}
\includegraphics[width=0.43\textwidth]{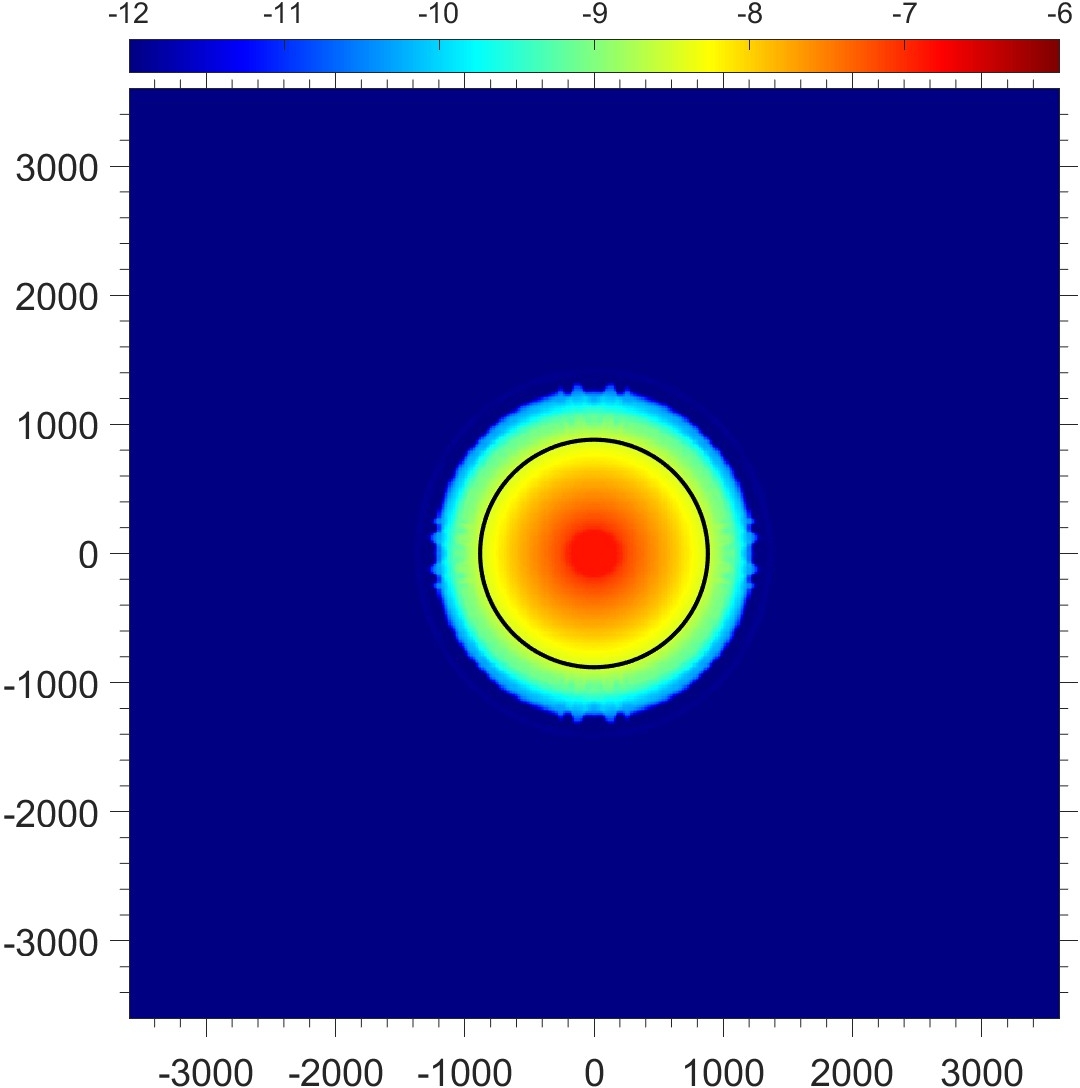}
\includegraphics[width=0.45\textwidth]{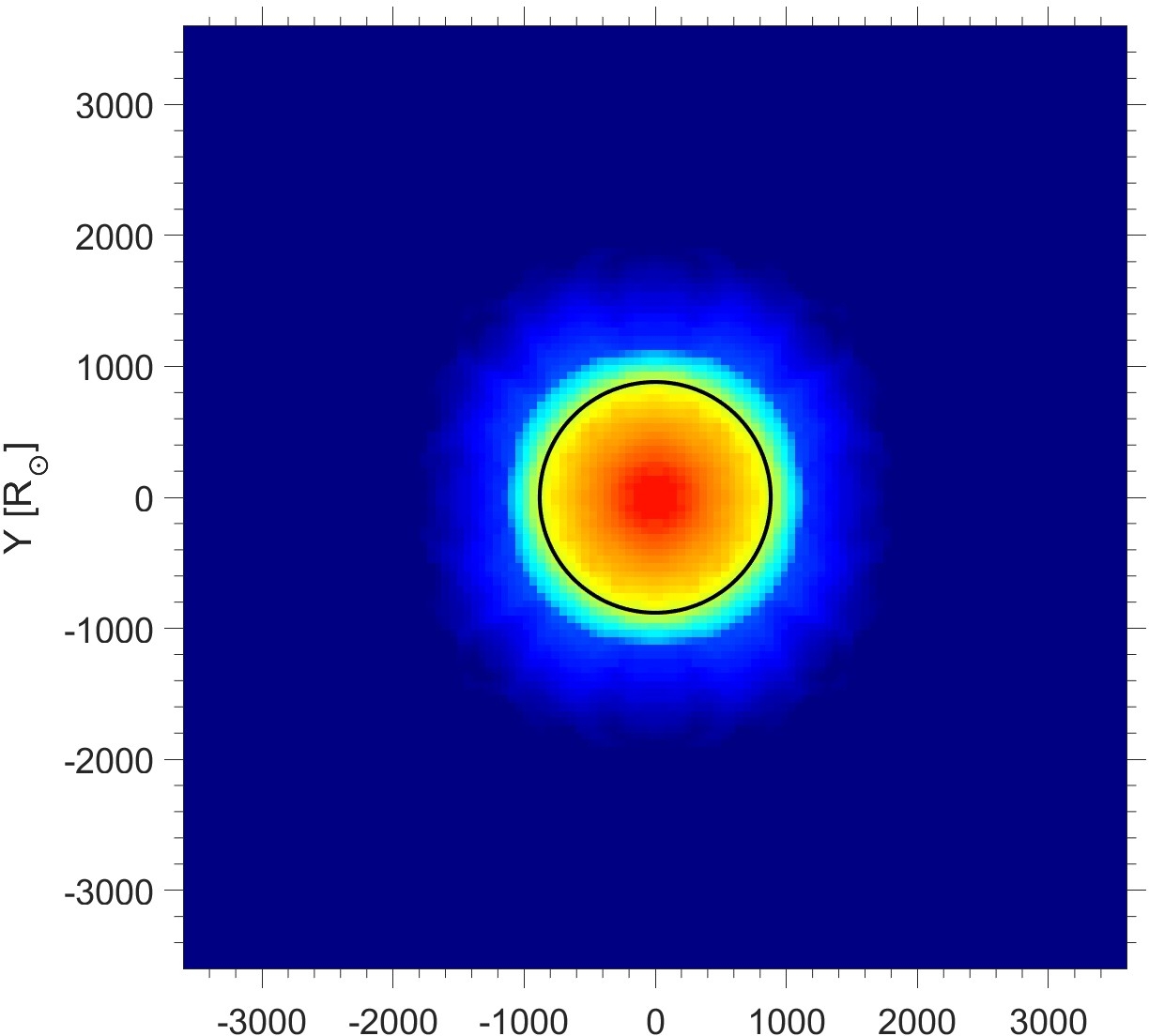}
\includegraphics[width=0.43\textwidth]{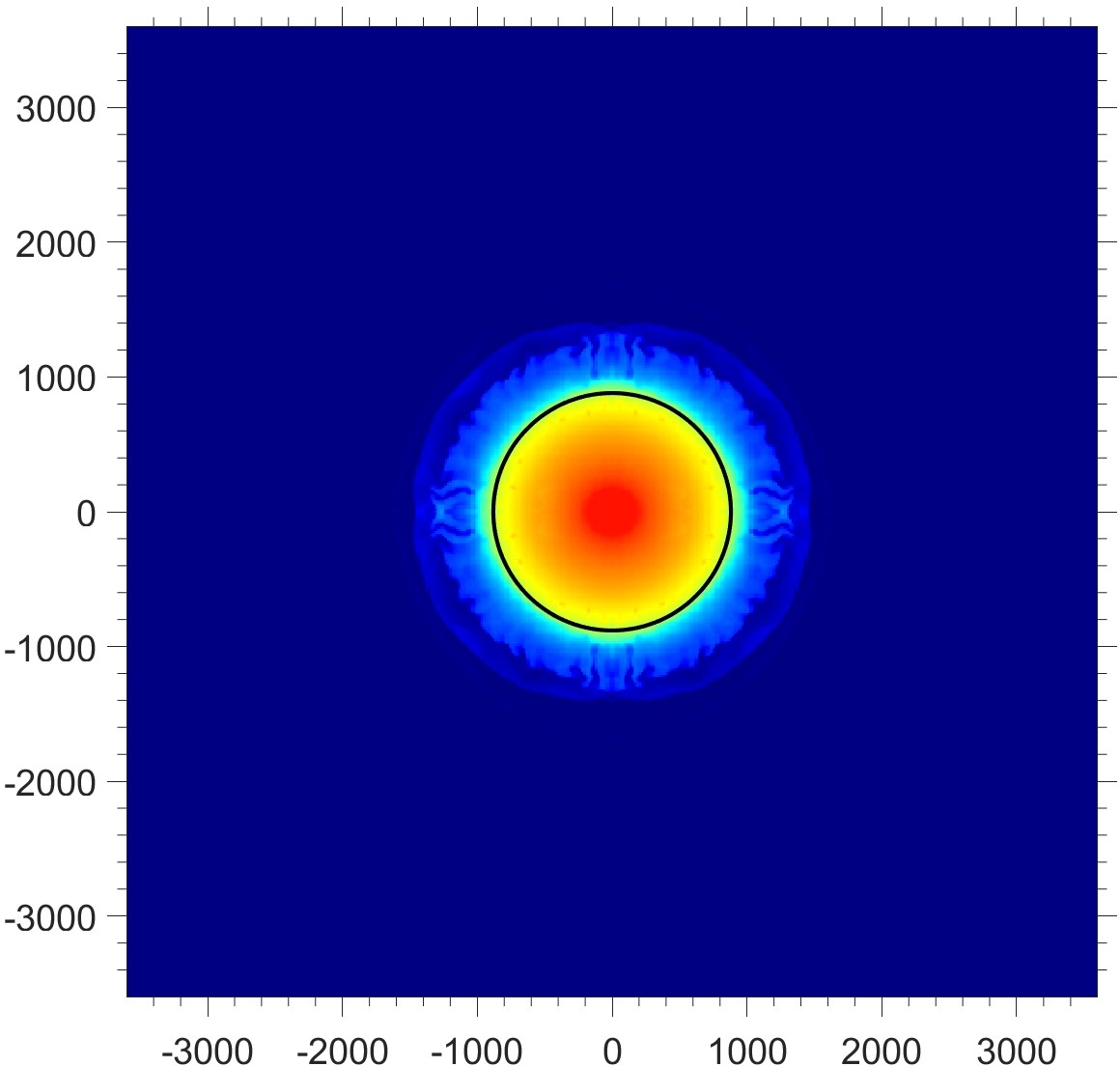}
\includegraphics[width=0.45\textwidth]{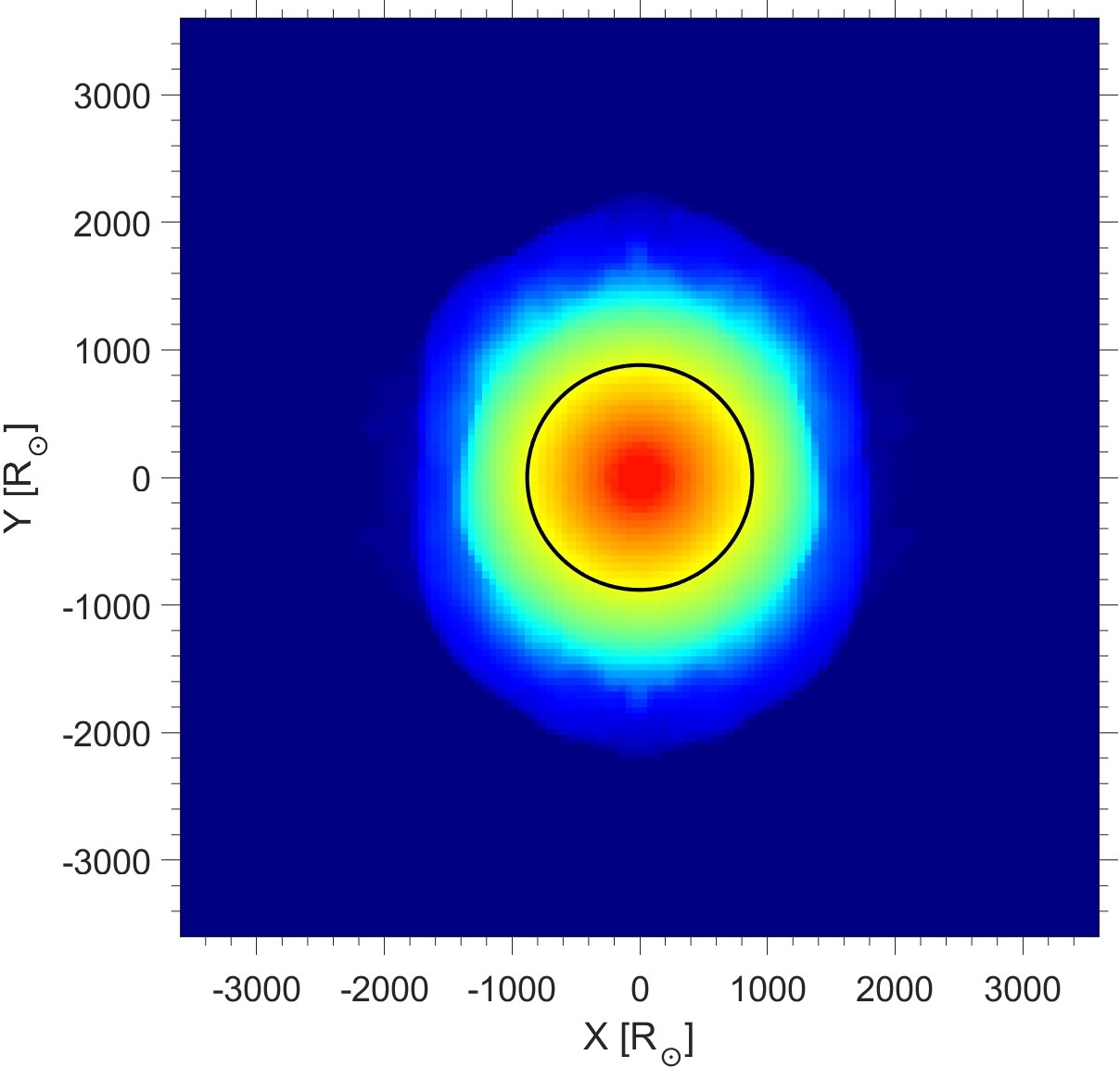}
\includegraphics[width=0.43\textwidth]{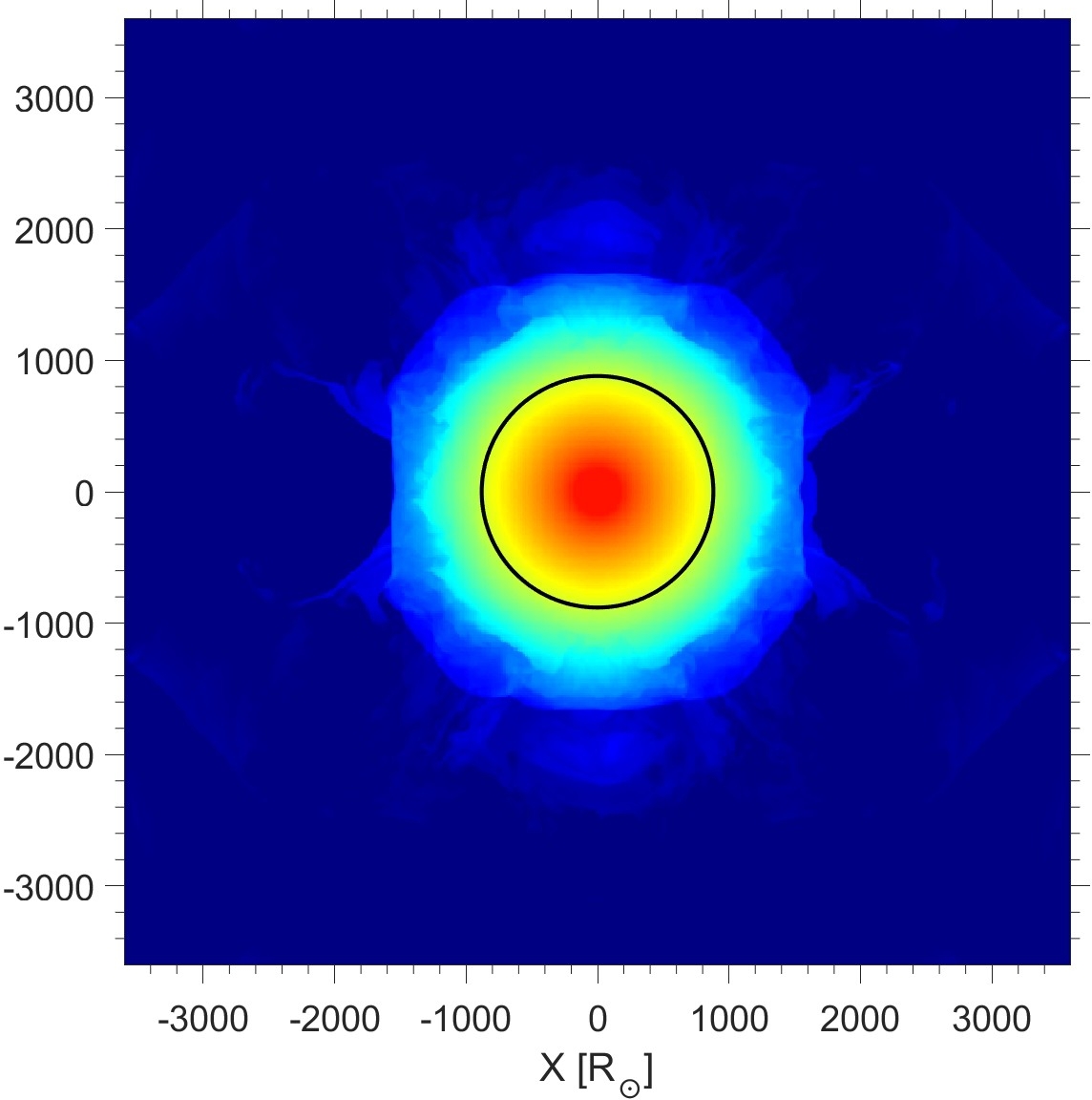}
\caption{Density maps of the regular-resolution (left column) 
and of the high-resolution (right column) simulations without any jets in the $z=0$ plane and at three times of, from top row to bottom, $t=0.7 \yr$, $t=1.6 \yr$ and $t=6.4 \yr$. 
The black circle in each panel marks the surface of the RSG model at $t=0$, which is $R_{\rm RSG}=881R_{\odot}=6.13 \times 10^{13} \cm$.
The density colour coding is according to the upper colour bar in logarithmic scale and in $\g \cm^{-3}$, from $10^{-12} \g \cm^{-3}$ (deep blue) to $10^{-6} \g \cm^{-3}$ (deep red).
}
\label{fig:DensityMaps}
\end{figure*}

In Fig. \ref{fig:DensityMaps} we present density maps in the plane $z=0$ for the regular (left column) and high (right column) resolution simulations without jets at three times (see caption). We mark the initial surface of the 1D (where the photosphere is well-defined) model with a black circle. The pale-blue and blue colours depict densities below the initial photospheric density, which is $\rho_{\rm p,0}= 2 \times 10^{-9} \g \cm^{-3}$. In these regions the results of our simulations are less reliable.  The general behaviour of the two resolutions is the same, but there are clear small-scale differences. 
The density maps reveal an important behaviour of the 3D stellar model, namely, that the star rapidly expands on a dynamical time scale of $\simeq P_{\rm D}$ and then contracts somewhat. It actually performs two oscillations before it relaxes.  

It is hard to follow the contraction from the density maps because we cannot follow the photosphere as we do not include radiative transfer. Instead, we mark each of two initial spherical shells within the star with `tracers', two different tracers for the two shells. We assign all cells inside a shell at $t=0$ a value of ${\rm tracer}=1$. As the material inside the initial shell mixes with gas outside the shell the value of the tracer decreases and it represents the fraction of mass in each cell that originated in the shell. The tracer value in each cell is always between zero and one. 
The two initial shells we follow are $600 R_\odot < r < 650 R_\odot$  and $800 R_\odot < r < 850 R_\odot$.

\begin{figure*} 
\centering
\includegraphics[width=0.43\textwidth]
{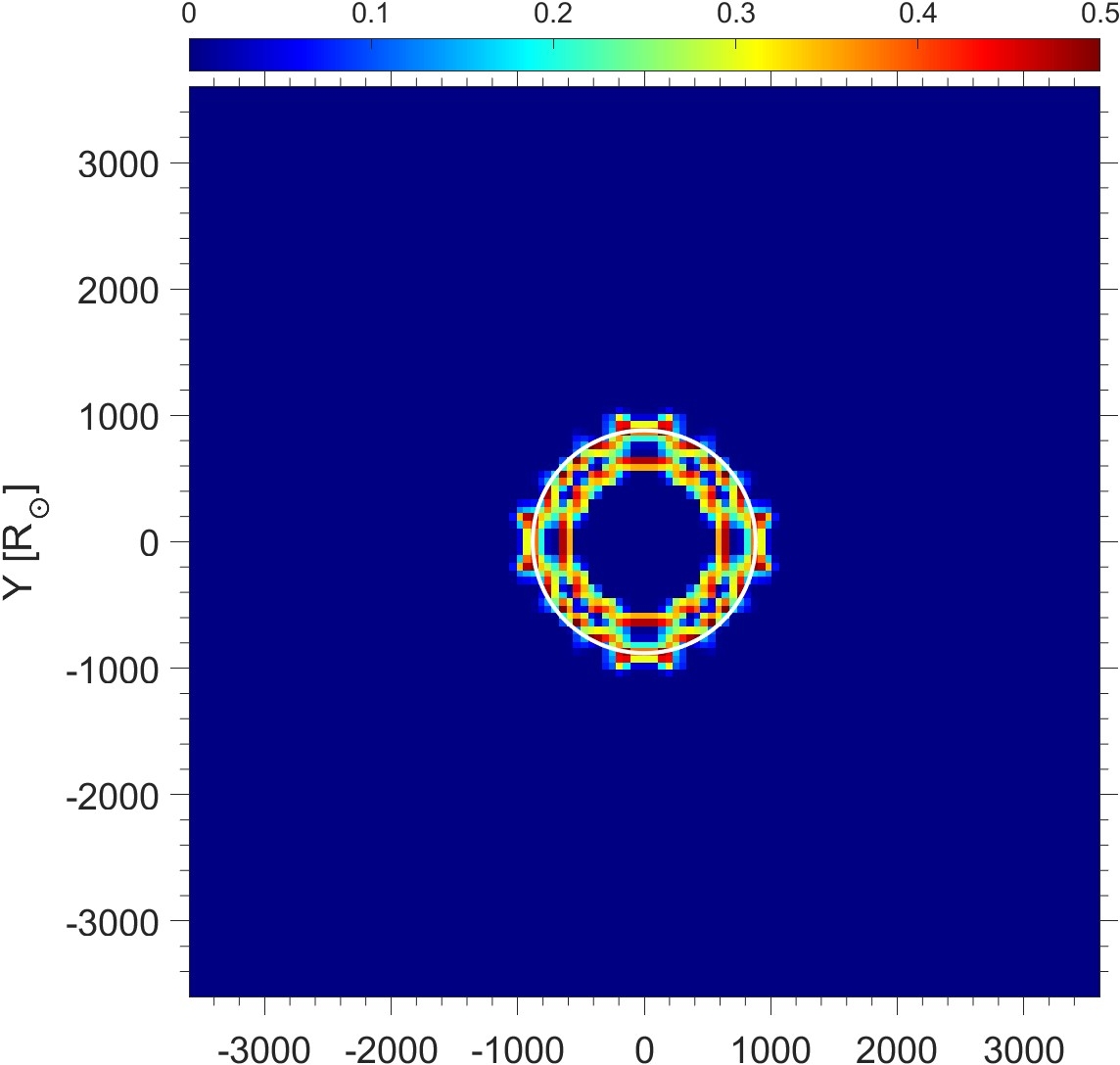}
\includegraphics[width=0.41\textwidth]
{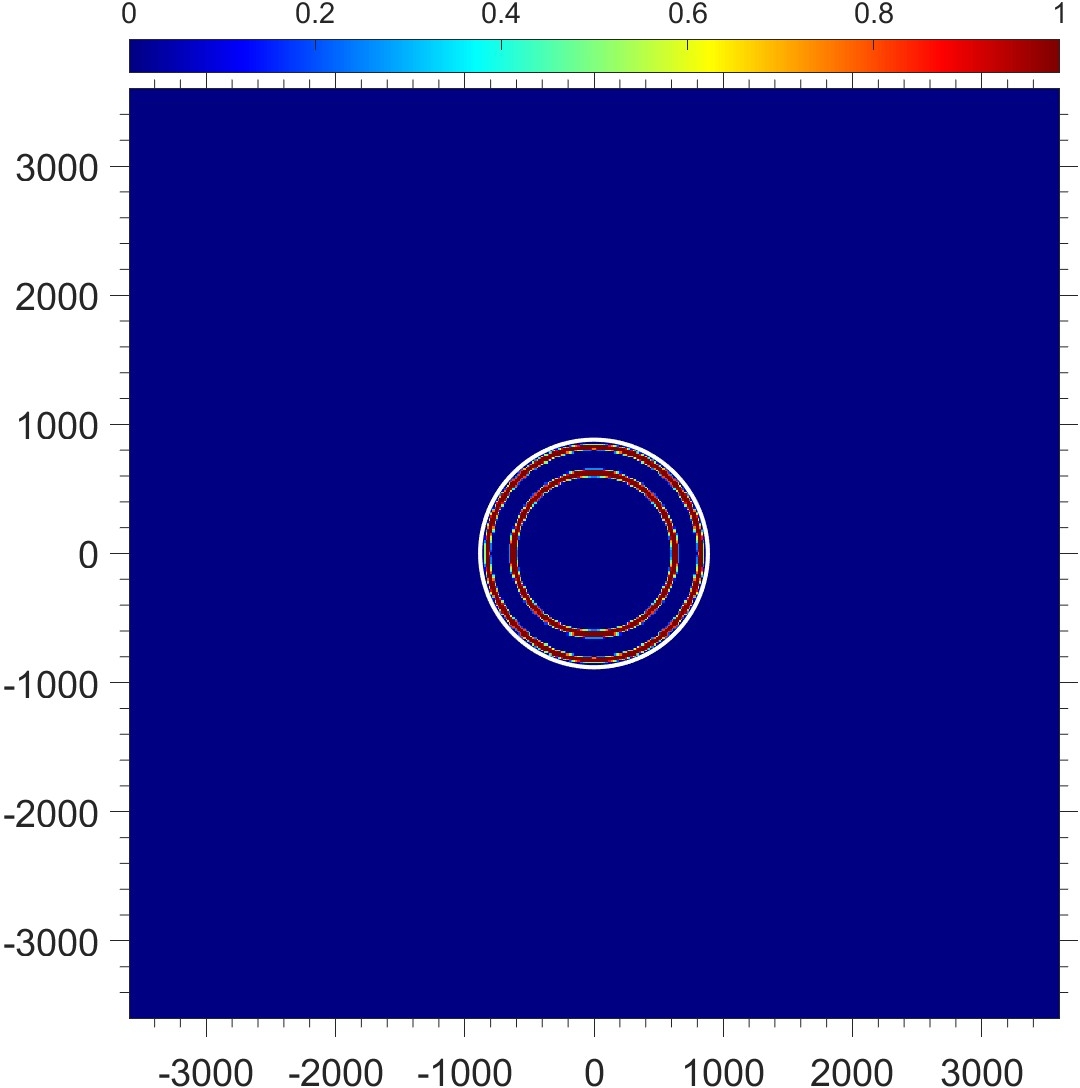}
\includegraphics[width=0.43\textwidth]
{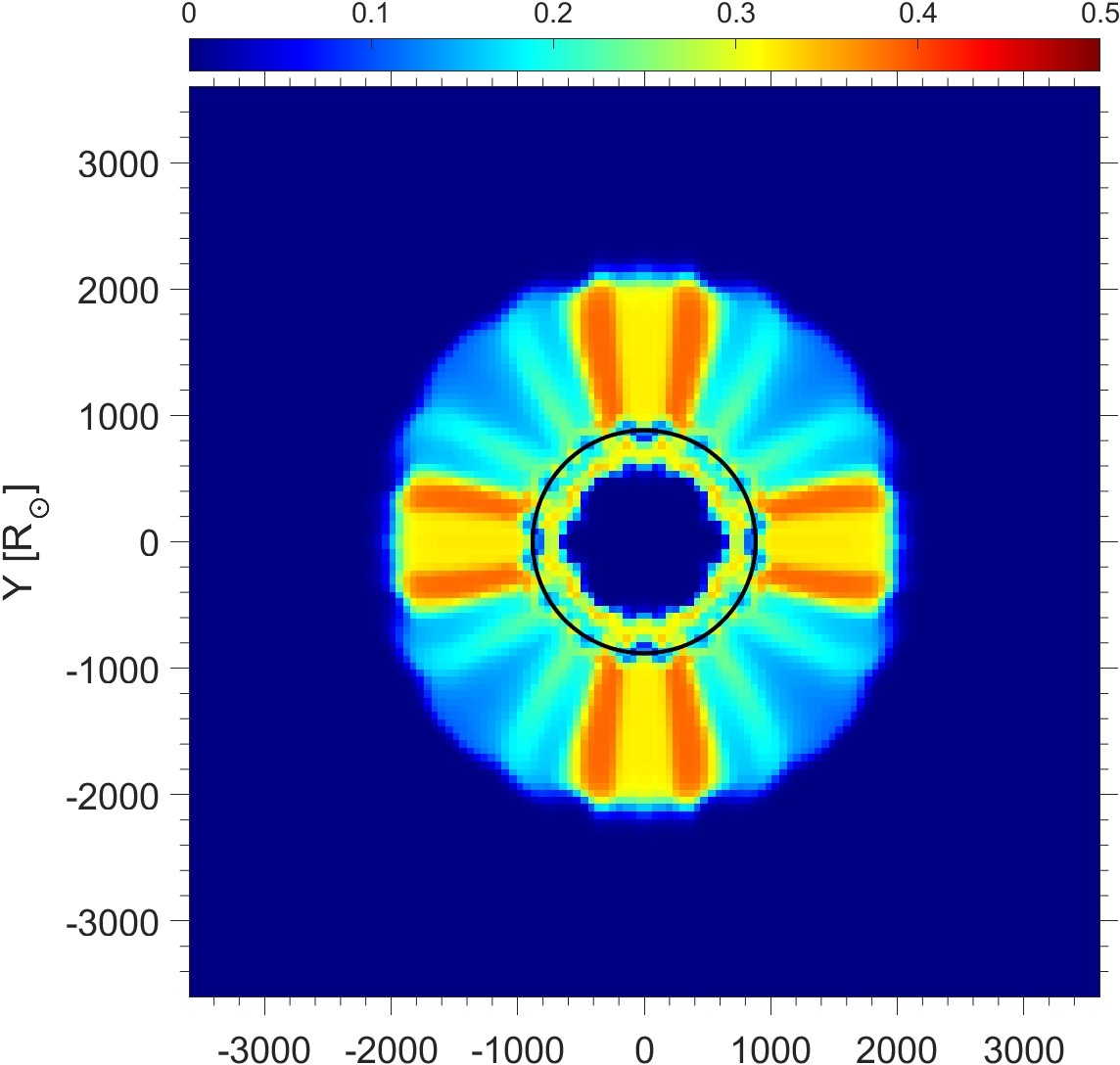}
\includegraphics[width=0.41\textwidth]
{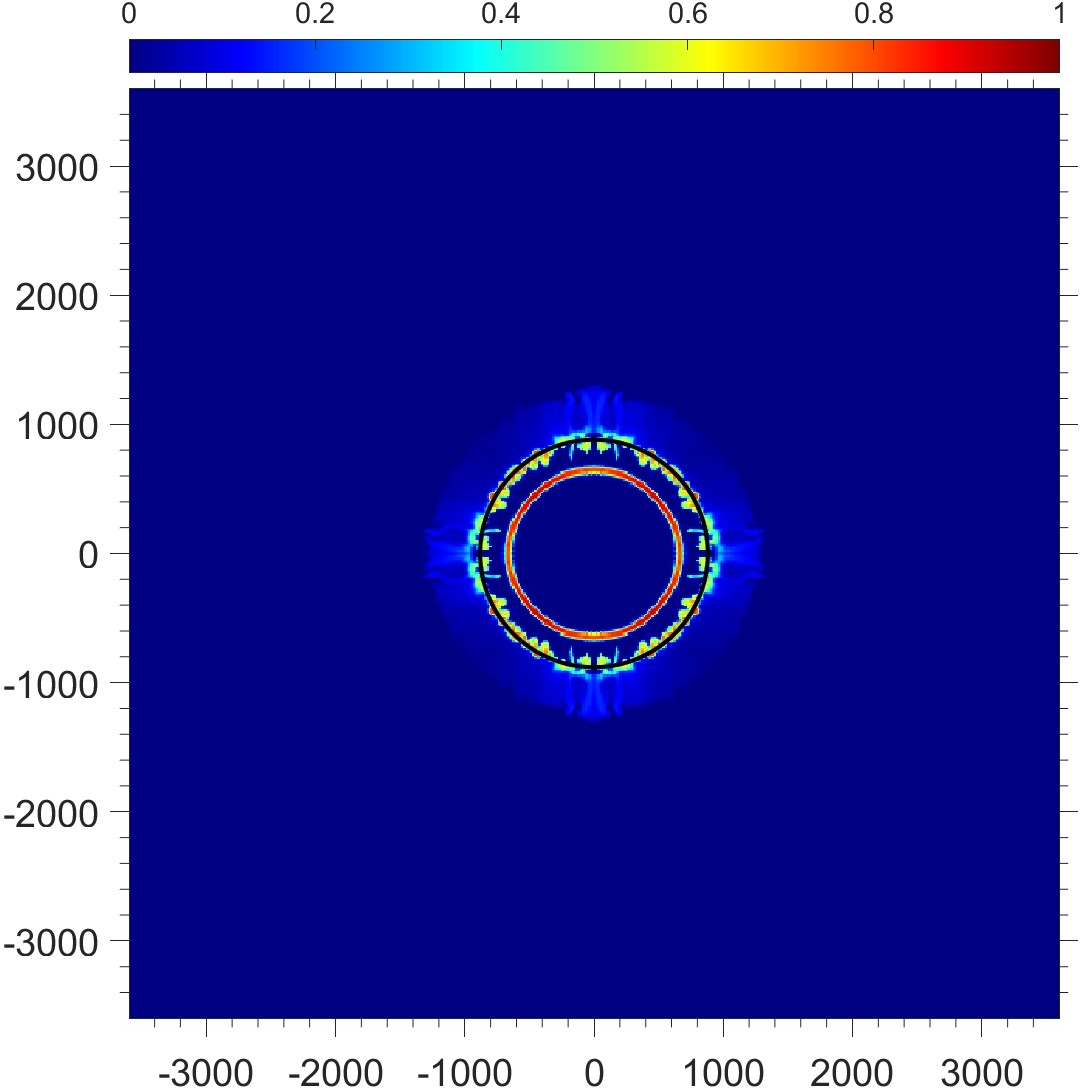}
\includegraphics[width=0.43\textwidth]
{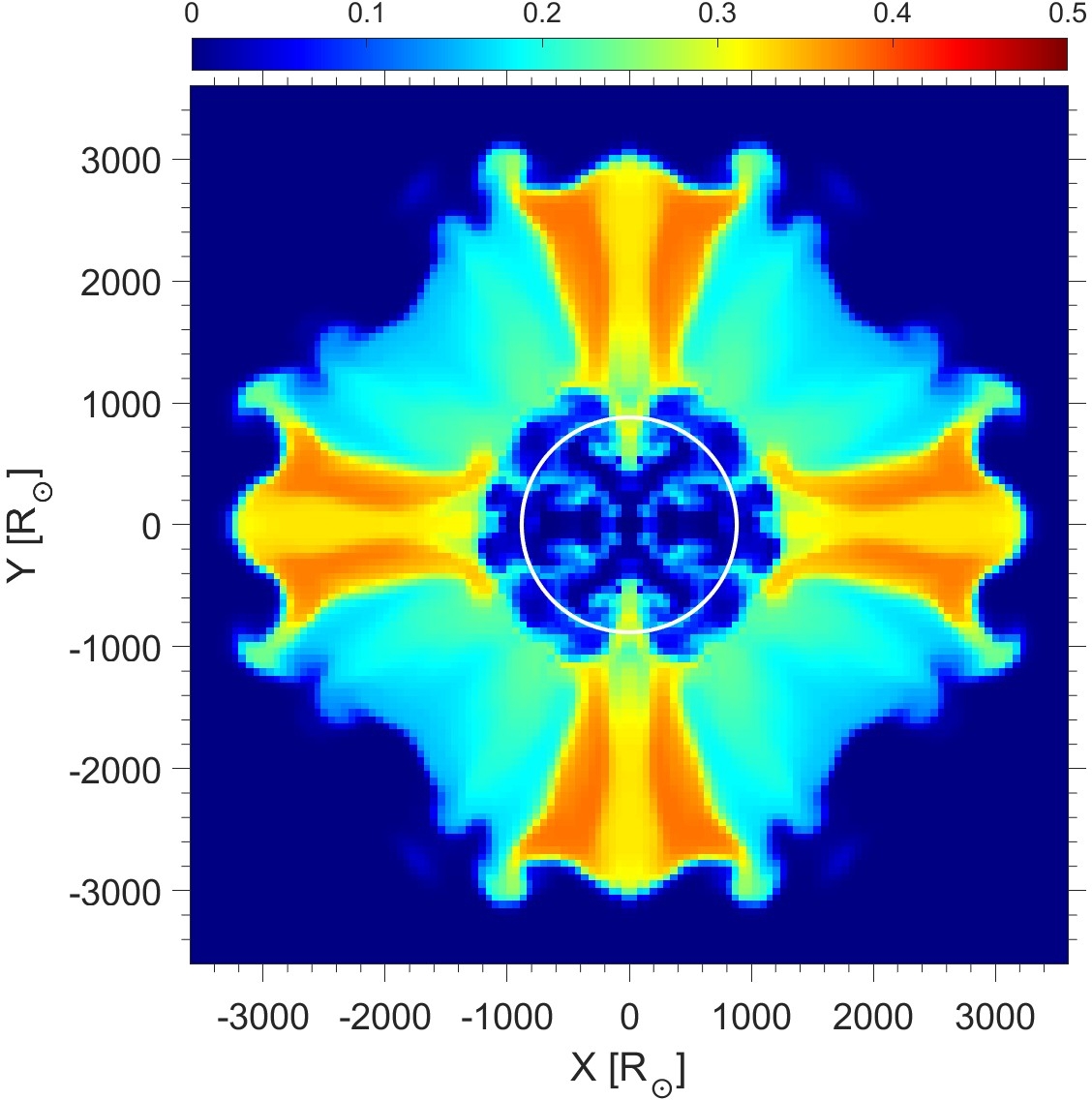}
\includegraphics[width=0.41\textwidth]
{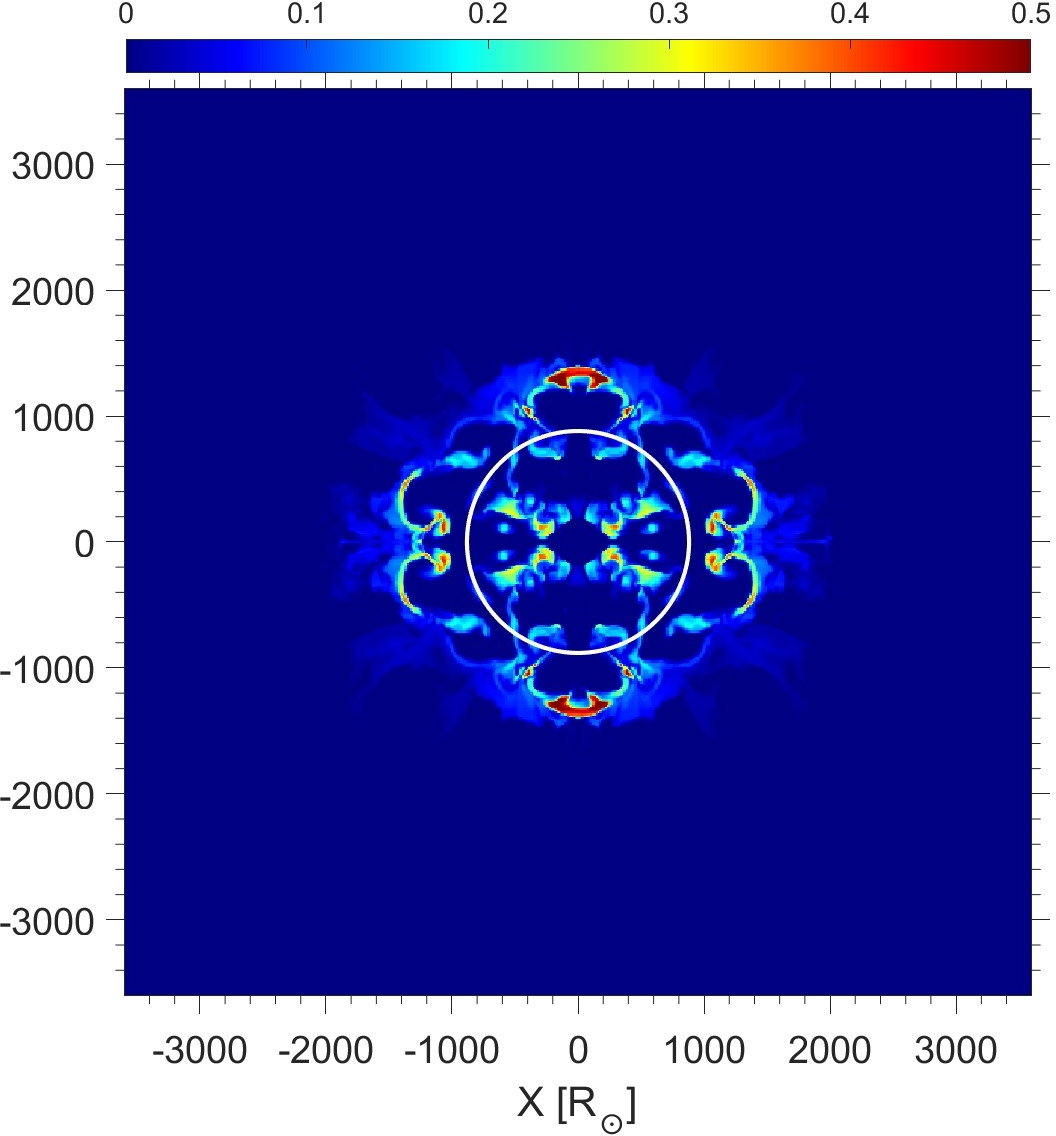}
\caption{Maps in the $z=0$ plane of the tracers of two shells (see text) in the regular resolution (left column) and in the high resolution (right column) of the no-jets simulations. 
The times from top to bottom are $t=0$, $t=1.5 \yr$ and $t=3.2 \yr$. 
The value of the tracer is according to the upper colour bar from $0$ (deep blue) to $1$ in the upper two panels in the right column, and to $0.5$ in the left column and in the lower right panel (deep red). The initial value of the tracer is $1$, but in the regular-resolution grid the cells do not fully resolve the initial shell.   
The circles with a radius of $R_{\rm RSG}=881R_\odot$ are the surface of the RSG star at $t=0$.
}
\label{fig:TracerMaps}
\end{figure*}

In Fig. \ref{fig:TracerMaps} we present the tracer maps at three times for the regular-resolution (left column) and the high-resolution  (right column) simulations. Here we clearly notice the limitation of the regular-resolution simulations. The cells in the regular-resolution cannot resolve the shells well, and the shells lose their identity very early in the simulations, i.e., in less than the dynamical time (not shown here) which suggests a large numerical effect. The high-resolution simulation maintains the identity of the shells for longer than the dynamical time. The mixing of the shells with each other and with the rest of the envelope that we see in the last panel on the right column (the high-resolution simulation) is a real physical effect due to the convection that is developed in the envelope (see below).   

\begin{figure} 
\centering
\includegraphics[width=0.44\textwidth]{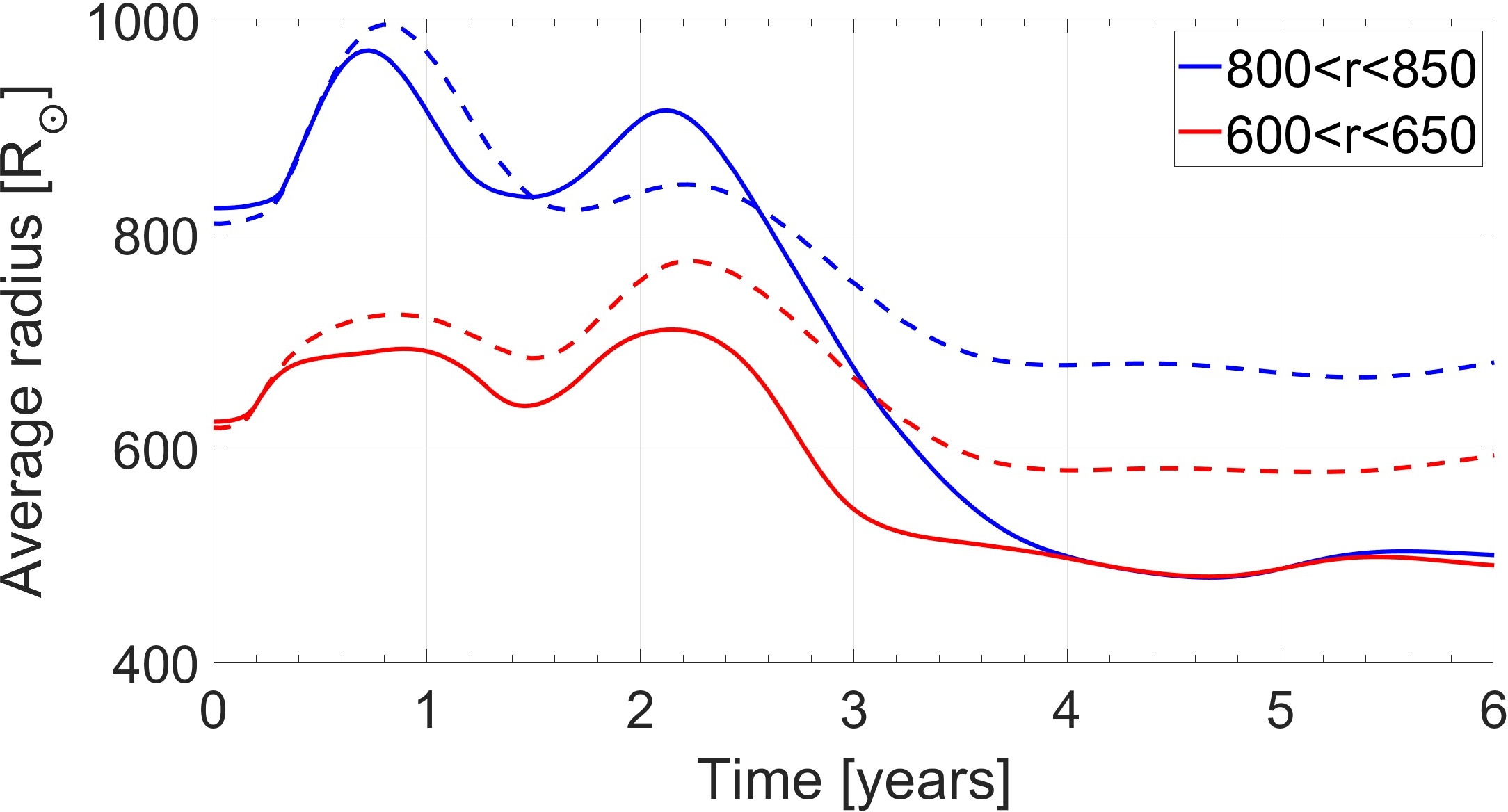}
\caption{The average radii of the tracers of two initial spherical shells $800 R_\odot < r < 850 R_\odot$ (blue lines) and $600 R_\odot < r < 650 R_\odot$ (red lines) in the simulations without jets.
The solid lines are for the high-resolution simulation and the dashed lines are for the regular-resolution simulation. For reference, the Keplerian orbital period on the surface of the unperturbed RSG is $P_{\rm Kep}=2.35  \yr$ and $P_{\rm D}\equiv (G \bar \rho)^{-1/2} =0.76  \yr$ where $\bar \rho$ is the average density of the initial RSG model. 
}
\label{fig:ShellsRadii}
\end{figure}

We also calculate at each time step the average radius of the gas that started in each of the initial two shells. We present the average radii of the gas that started in the two shells as a function of time in Fig. \ref{fig:ShellsRadii}.  

Figs. \ref{fig:TracerMaps} and \ref{fig:ShellsRadii} show two prominent types of behaviour. (1) The two shells together with the entire star perform two oscillations (two maxima) before the star relaxes. (2) The two shells in the high-resolution simulation are mixed with each other on a time scale of $\simeq 3.4 \yr \simeq 1.5 P_{\rm Kep}$. The two shells in the regular-resolution simulation maintain their average separation but are mixed as well. We follow only the tracers of the two shells, but the entire envelope is actually mixing with itself. Both of these types of behaviour are physically real, as we now discuss. 

Long period variables (LPVs) reach, in the non-linear regime, variations in their radius (maximum radius minus minimum radius in a cycle) that are about equal to their average radius, $\Delta R \simeq R$ (e.g., \citealt{Trabucchietal2021} for a recent study). With our code we cannot follow the photosphere as we include no radiative transfer. If we take the average radius of the outer shell that we follow (blue lines in Fig. \ref{fig:ShellsRadii}) before the shells are smeared, i.e., $t < 3 \yr$ (see Fig. \ref{fig:TracerMaps}), we find the average radius to be $\bar R \simeq 850 R_\odot$ and the variation to be $\simeq 200 R_\odot$. The photosphere is somewhat larger than the average radius of the shell. Examining the sharp edge of the models in the first two panels of Fig. \ref{fig:ShellsRadii} we find $\Delta R/\bar R \simeq 0.25-0.3$. 
Comparing the variations in the RSG radius during the time $t \la P_{\rm Kep} \simeq 3 P_{\rm D}$ with the theoretical results of, e.g., \cite{Trabucchietal2021}, we conclude that the 3D stellar model simply performs the expected oscillations of RSG stars, likely in the non-linear regime. 

Consider then the mixing of the two shells, and actually the entire envelope. 
From the 1D model we know that the envelope of our model is unstable to convection, i.e., entropy decreases outward. What we find here is that the initial static 3D model develops convection that flattens the entropy profile. 

\begin{figure*} 
\centering
\includegraphics[width=0.45\textwidth]
{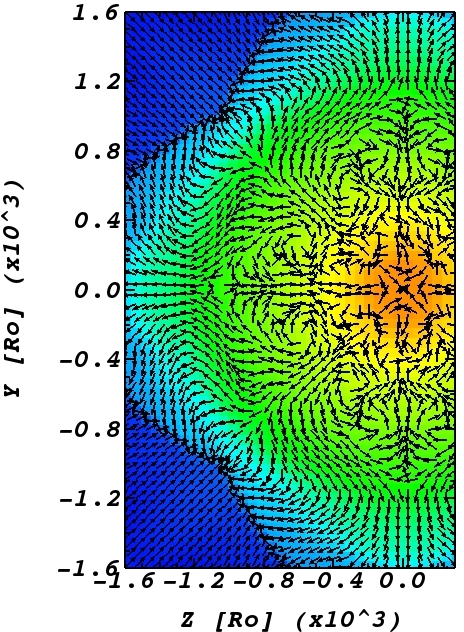}
\includegraphics[width=0.45\textwidth]
{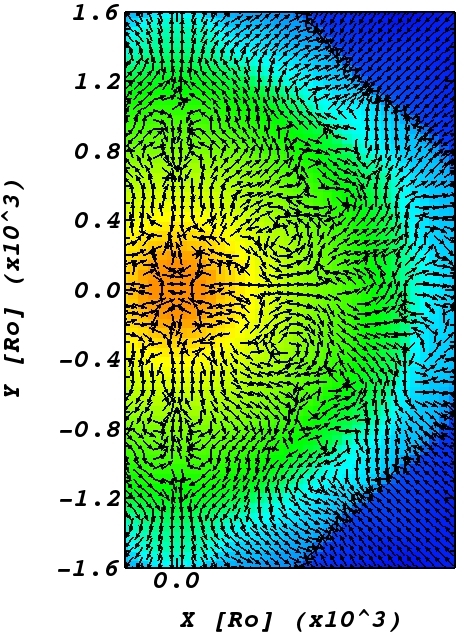}\\
\includegraphics[width=0.45\textwidth]
{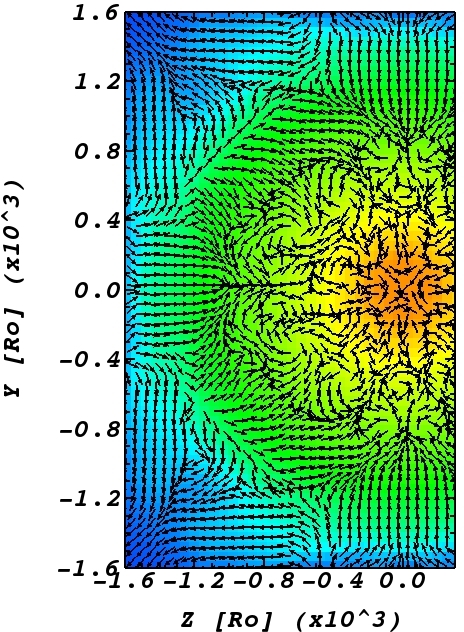}
\includegraphics[width=0.45\textwidth]
{Den_vel_vec_XY_j124_LR_new.jpg} \\
\caption{Velocity vectors on top of density maps 
in part of the $x=0$ plane (left) and in part of the the $z=0$ plane (right), at $t=4 \yr$ (top), and $t=4.5 \yr$ (bottom) for the regular-resolution no-jets simulation. The density scale and units of axes are as in Fig. \ref{fig:DensityMaps}. The flow speed at each point is according to the length of the arrow. The maximum velocities in the vortices at $700 R_\odot \simeq 50 \times 10^{12} \cm$ are $v_{\rm con,3D} \simeq 3 \km \s^{-1} - 8 \km \s^{-1}$.    
}
\label{fig:VorticiesCuts_LR}
\end{figure*}

\begin{figure*} 
\centering
\includegraphics[width=0.45\textwidth]
{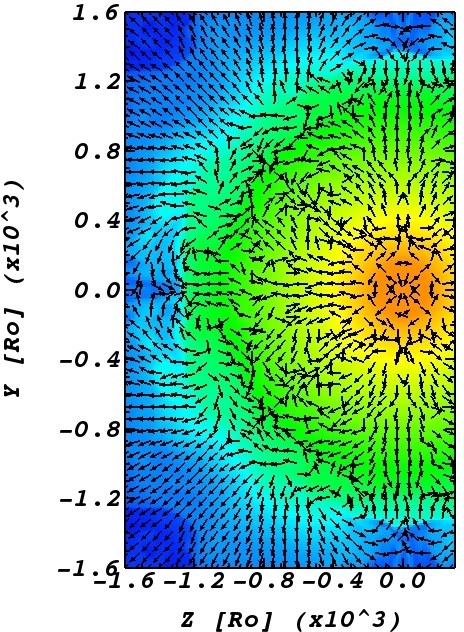}
\includegraphics[width=0.45\textwidth]
{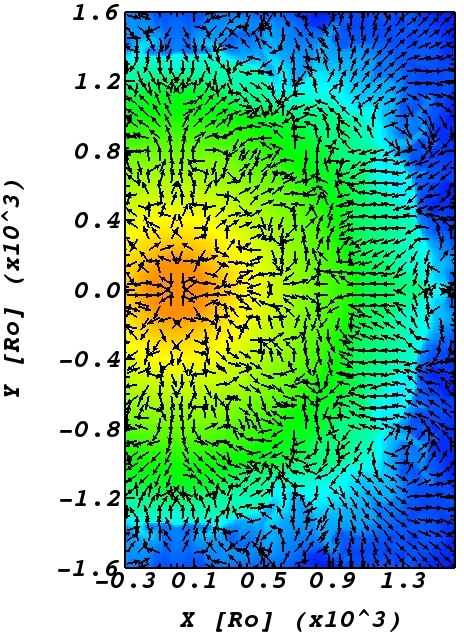} \\
\includegraphics[width=0.45\textwidth]
{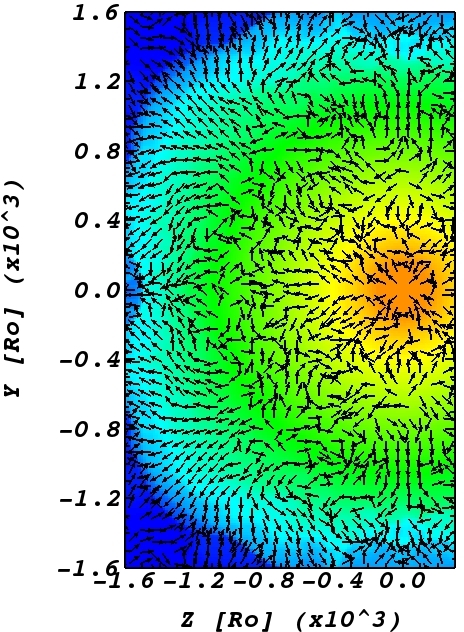}
\includegraphics[width=0.45\textwidth]
{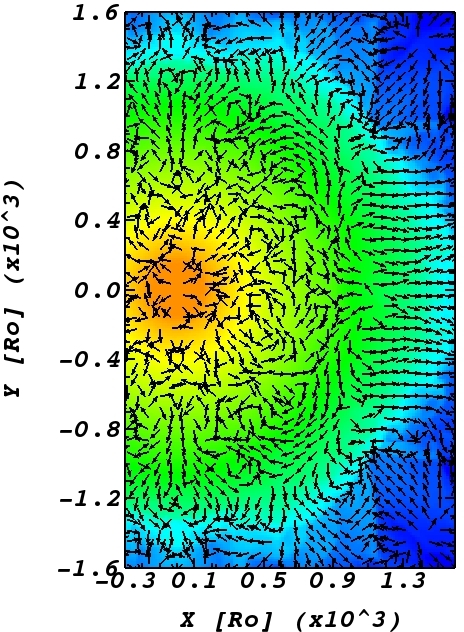} \\
\vspace{8mm}
\caption{Similar to Fig. \ref{fig:VorticiesCuts_LR} but for the high-resolution simulation. 
}
\label{fig:VorticiesCuts_HR}
\end{figure*}

We first present the velocity maps in two planes and at two times, separated by about the dynamical time, in Fig. \ref{fig:VorticiesCuts_LR} for the regular-resolution simulation and in Fig. \ref{fig:VorticiesCuts_HR} for the high-resolution simulation. As said, we do not study here the flow in the very-low density zones (pale-blue and blue regions) because of large numerical uncertainties. The red inner zone is the inert core and we do not consider the flow in its vicinity. We discuss the flow structure in the green and yellow zones.    

We note the following flow properties. 
(1) There are large vortices as obtained in other 3D simulations of convection (e.g., \citealt{GilkisSoker2016, FieldsCouch2020}). 
(2) The velocity is stochastic. We infer this property by comparing the velocity maps in the same planes at the two times $t=4 \yr$ and $t=4.5 \yr$ in Figs.  \ref{fig:VorticiesCuts_LR} and Fig. \ref{fig:VorticiesCuts_HR}. The velocity structure substantially changes between these two times separated by about the dynamical time. We also learn this from the different flow structures in the two perpendicular planes at a given time. 
(3) We find that the typical maximum convective velocity in our simulations in the shell around $r \simeq 700 R_\odot$ 
is $v_{\rm con,3D} \simeq 3 \km \s^{-1} - 8 \km \s^{-1}$. 
In the 1D model that we used to build the 3D model the typical convective velocity, from the mixing length theory, at the same zone is $v_{\rm con,1D}=5 \km \s^{-1}$. We find that $v_{\rm con,3D} \simeq v_{\rm con,1D}$. However, we do not include the nuclear energy source which would have forced the convection speed to be larger. More accurate simulations find the 3D convective velocity to be larger. 
\cite{FieldsCouch2021} find that the angle-averaged convective speeds in the core of massive stars before core collapse are 3-4 times larger than the values of 1D models by \textsc{MESA}.  

\begin{figure} 
\centering
\includegraphics[width=0.40\textwidth]{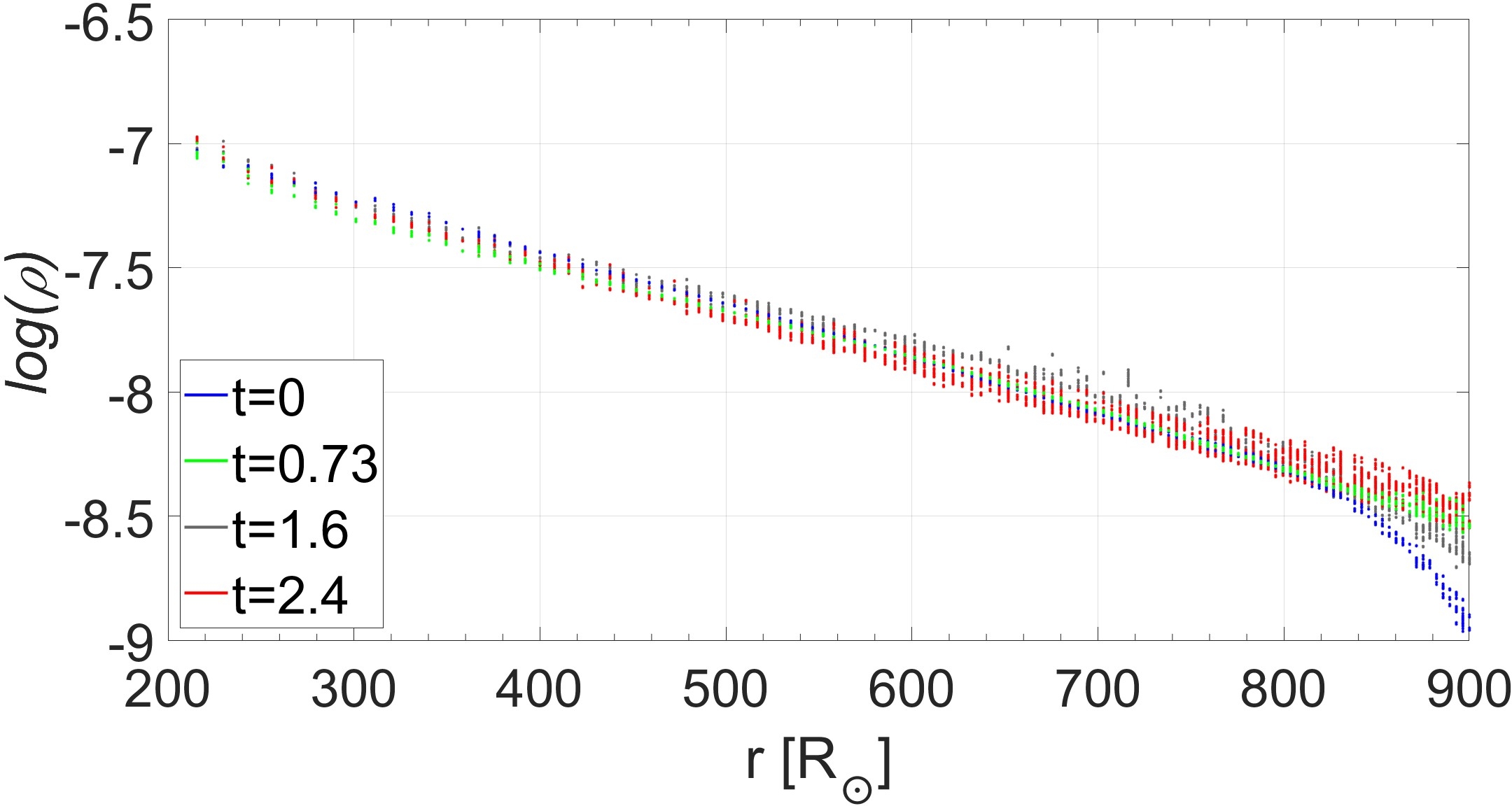}
\includegraphics[width=0.40\textwidth]{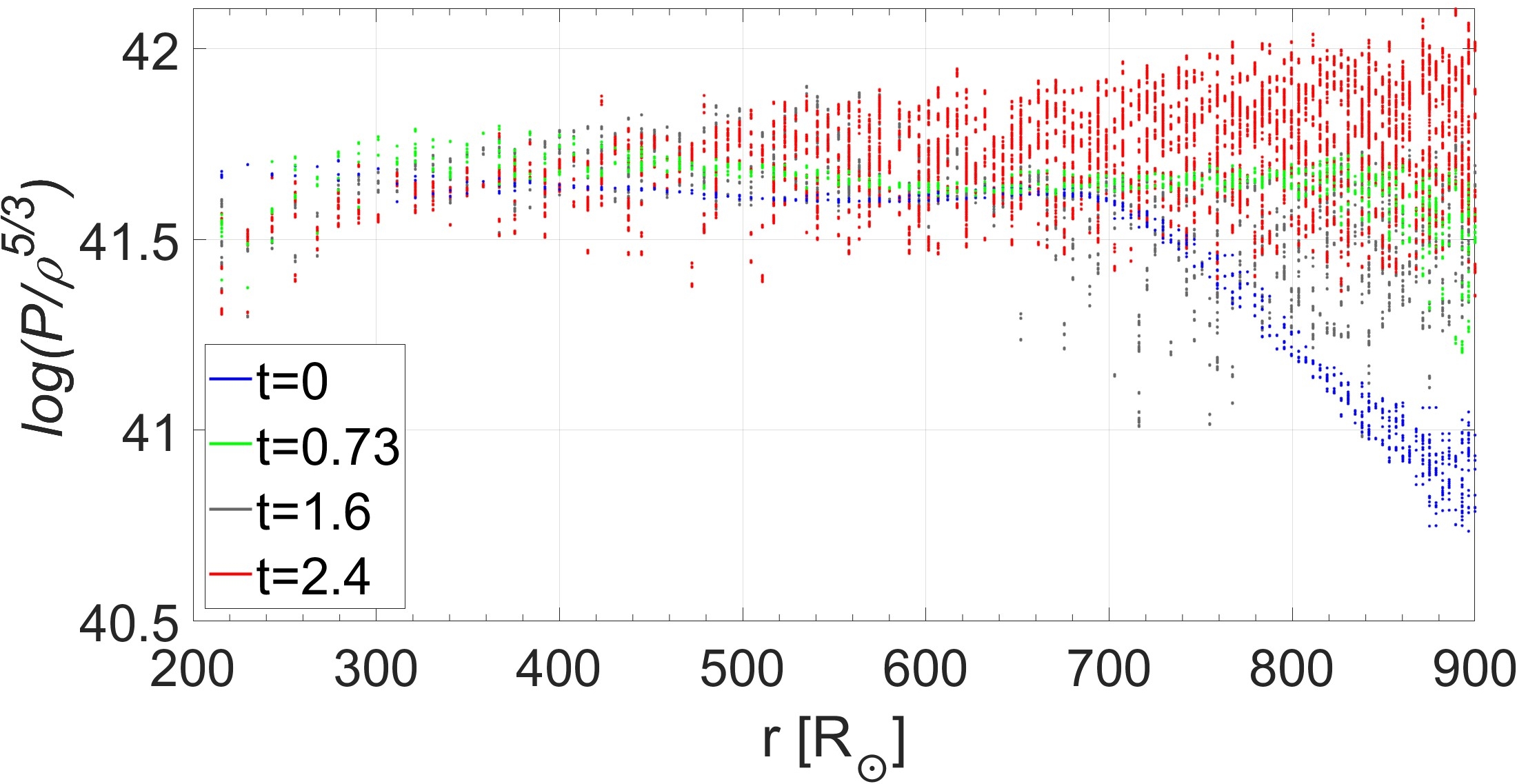}
\caption{Radial profiles of the density (top; density in $\g \cm^{-3}$)
and of $F_{\rm S} = log{(P/\rho^{5/3})}$, which is about proportional to the entropy (bottom; $P/\rho^{5/3}$ in units of $\g^{-2/3} \cm^{4} \s^{-2}$) at four times as indicated in the insets and for the regular-resolution no-jets simulation.
Each scatter-plot is the collection of the values in all cells at the given time (a total of about $43,000$ points). 
}
\label{fig:Entropy_LR}
\end{figure}

\begin{figure} 
\centering
\includegraphics[width=0.40\textwidth]{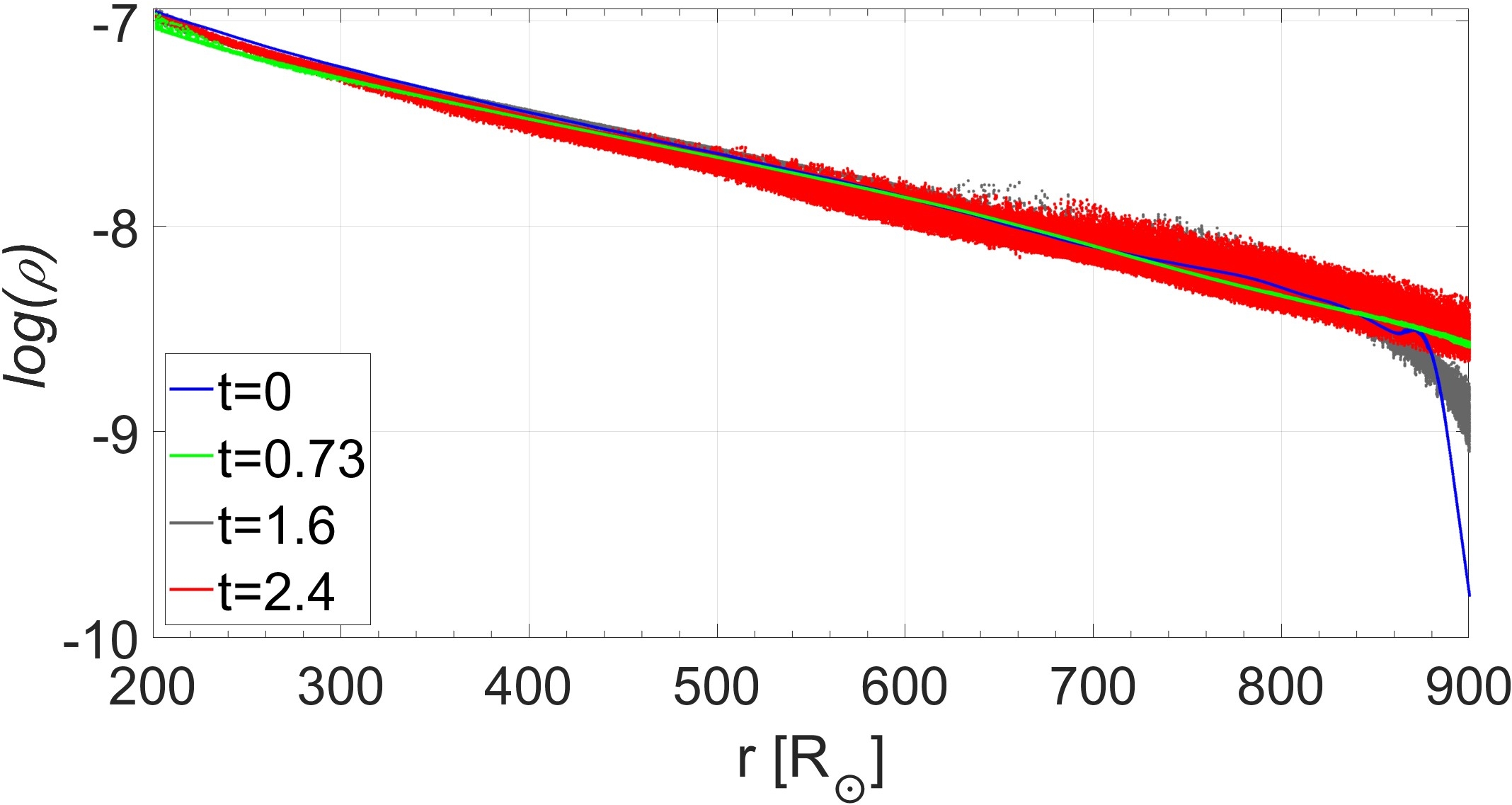}
\includegraphics[width=0.40\textwidth]{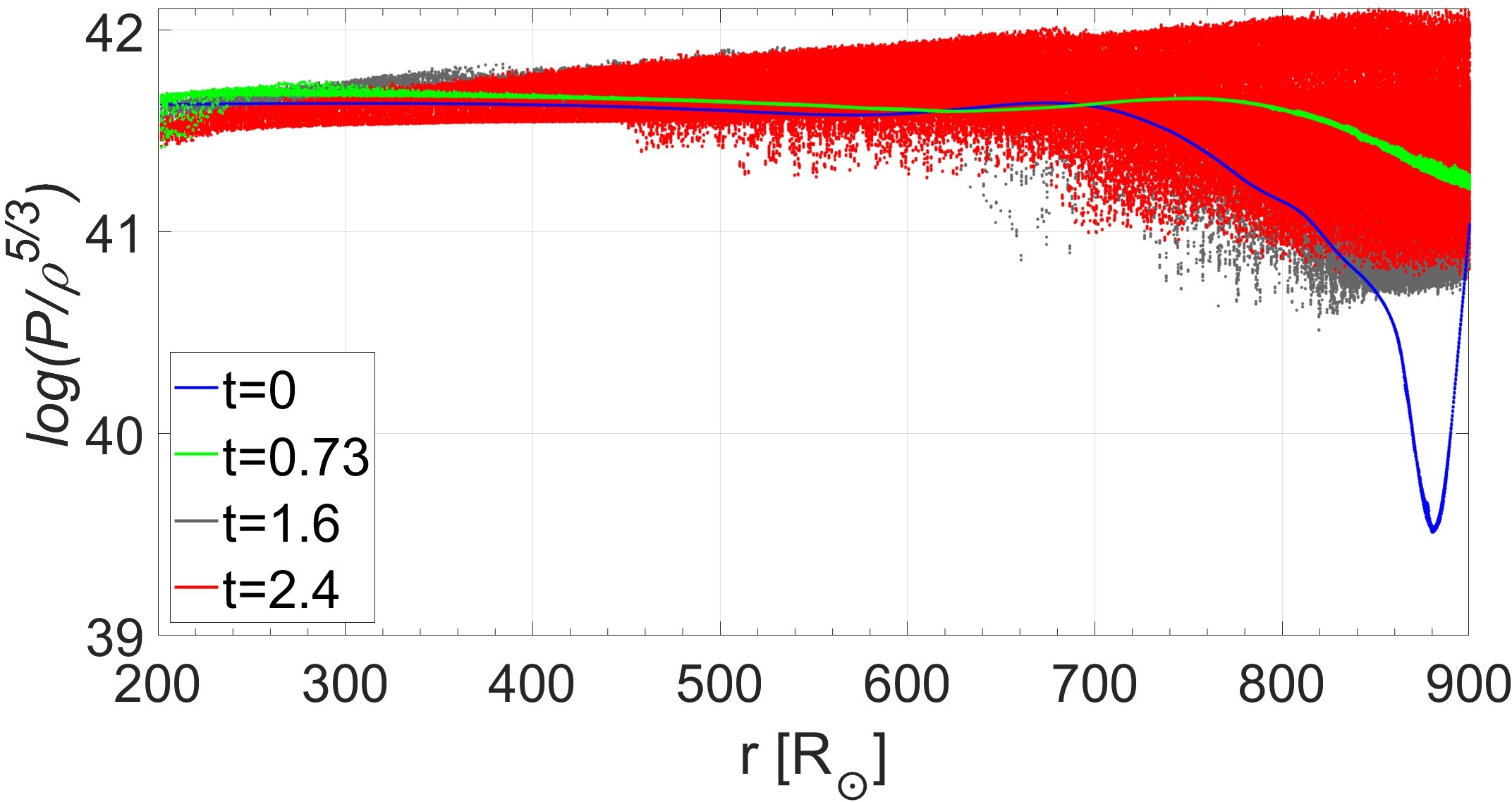}
\caption{Similar to Fig. \ref{fig:Entropy_LR} but for the
high-resolution simulation where we have $1.1\times10^6$ points in each scatter-plot. 
}
\label{fig:Entropy_HR}
\end{figure}

In Figs. \ref{fig:Entropy_LR} and \ref{fig:Entropy_HR} we present the evolution of the density profile (in units of $\g \cm^{-3}$) and of the profile of the quantity $F_{\rm S} \equiv \log (P/\rho^{5/3})$ (in units of $\g^{-2/3}\cm^{4} \s^{-2}$) which is about proportional to the entropy (the accurate entropy profiles have very similar slopes, but this function is easy to follow here). We present the profiles at four times as indicated. 
The plots are made of points, each representing one cell in the numerical grid. The initial slope of $F_{\rm S}$ has large regions of negative gradient $dF_{\rm S}/dr <0$ which implies convectively unstable zones. And indeed, convection sets in and flattens the entropy profiles (green to black to red colour). The density profile does not change much in most of the envelope. Only in the outer parts the density increases.  

We found that the regular-resolution simulations add mass above the envelope due to the steep density gradient. This makes the results of the no-jets simulation unreliable in those outer regions at times of $t \ga P_{\rm Kep} \simeq 2.5 \yr$. The high-resolution simulations is reliable up to $t \simeq 3 P_{\rm Kep}$ in the outer regions and for much longer times in the inner regions. In the simulation with jets we are forced to use the regular-resolution grid. Because in a short time the NS spirals in (section \ref{sec:Turbulence}) and  the inner regions are reliable, we find that our regular-resolution grid is adequate for the purposes of the present study.   

We summarise this section as follows. The implication of our results to numerical simulations that involve giant stars (RGB, AGB, RSG) that perform binary interaction is that there is no need to stabilise the giant star to a high degree when transferring a 1D model to a 3D grid. 
If the 3D stellar model performs oscillations, even non-periodic oscillations in the non-linear regime ($\Delta R \approx R$), this is perfectly fine as such stars are expected to have large-amplitude non-regular pulsations. In addition, instabilities as we present in Fig. \ref{fig:TracerMaps} are expected in the convective envelope of giant stars.  
Researchers simulating interacting red giant stars should just let the 3D model oscillate in large amplitudes and mix different layers by turbulence. These processes are more realistic than building a completely stable red giant model.  

\section{Jet-powered turbulence} 
\label{sec:Turbulence}

In this section we set jets from an orbiting NS. We do not include the mass or the gravity of the NS, but rather only the jets we assume the NS launches perpendicular to the orbital plane. As we described in section \ref{subsec:StellarOrbit} 
we let the NS to spiral-in along a predetermined orbit from $a_{\rm i} = 850\,R_{\odot}$  to $a_{\rm SR} = 300\,R_{\odot}$ in a time of 3 years. It then performs a circular orbit at radius $a_{\rm SR}$.  The radial velocity during the rapid in-spiral (plunge-in) phase is constant.
The power of the two jets changes according to equation (\ref{eq:JetsPower}) with $\zeta =2\times 10^{-5}$ along the entire evolution. We find that the power of the jets increases from $\dot E_{2j} (a_{\rm i}) = 3.3\times 10^{40} \erg \s^{-1}$ at the outer orbital separation to $3\times 10^{41} \erg \s^{-1}$ at the final orbit $a_{\rm f}$. 

\begin{figure} 
\centering
\includegraphics[width=0.45\textwidth]{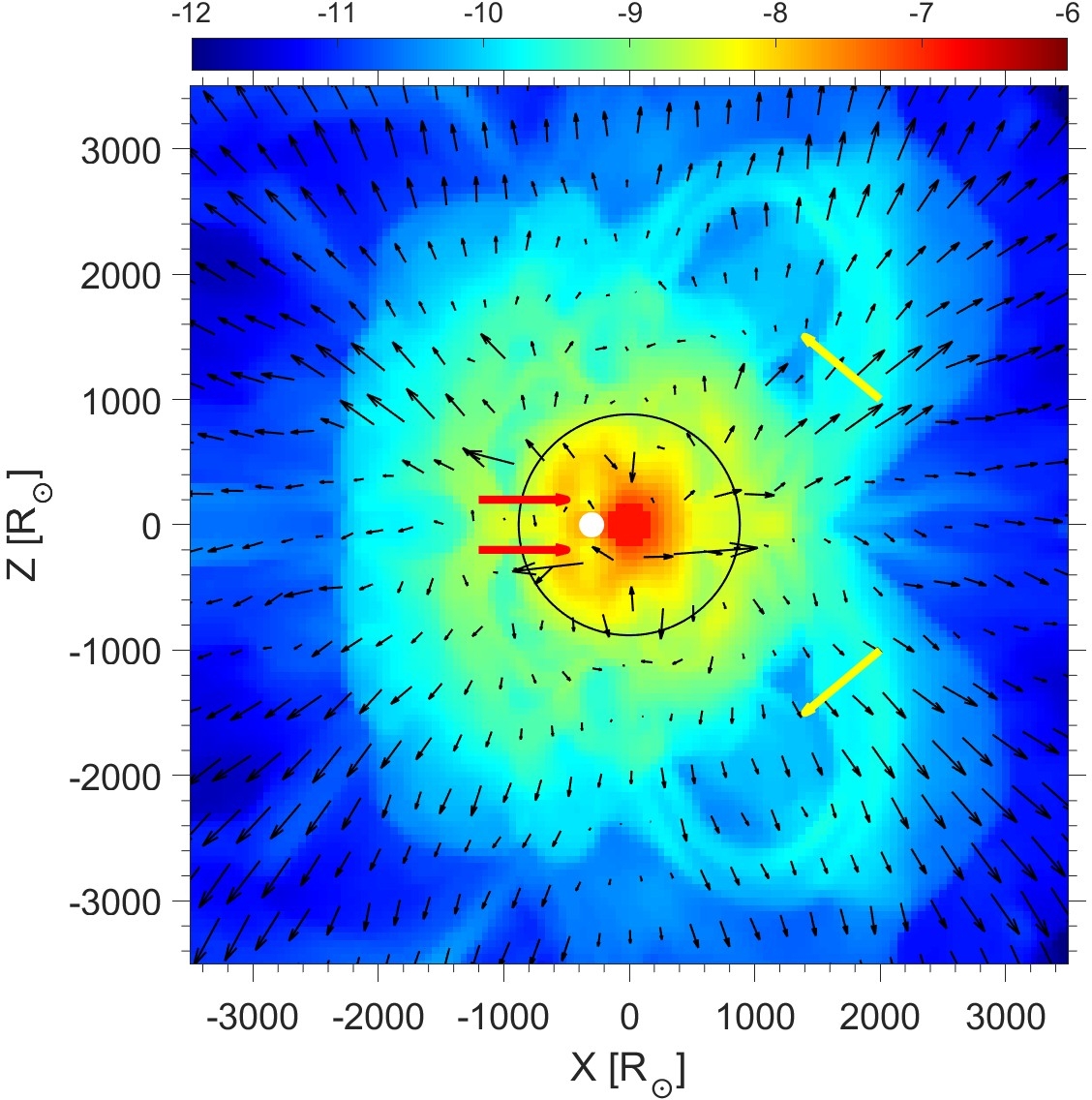}
\caption{ Velocity arrows on top of the density maps of the spiralling-in simulation with jets in the $y=0$ plane at $t=3.2 \yr$, when the NS orbits at $a_{\rm SR}=300 R_\odot$. This plane cuts the low-density and high-outflow velocity bubbles that the jets inflate in two regions above and two regions below the equatorial plane. The older two regions (above and below the equatorial plane) are to the right of the centre and appear as pale-blue zones surrounded by green regions;  we point at these regions of the bubbles with yellow arrows. The two newer regions are to the left of the centre and appear as two green zones surrounded by yellow (higher density) regions;  we point at these regions of the bubbles with red arrows. Velocity magnitude is proportional to the length of the arrows. Maximum velocity inside the old bubbles is $76 \km \s^{-1}$.  At the time of this figure the NS is at $(x,y,z)_{\rm NS}=(-151 R_\odot, 259 R_\odot, 0)$, i.e., outside the plane of the figure. The white dot is the projected location of the NS on the $y=0$ plane. 
}
\label{fig:rho_SPIRAL6}
\end{figure}

The jets that the NS launches induce fast flows inside the envelope, including regions of outflows and regions of vortices.  The jets inflate two bubbles at mid-latitudes in the expanding envelope. In 3D the morphology of each bubble is a low-density spiral that expands outward along mid-latitude directions. We can see the cross section of these two bubbles, one above and one below the equatorail plane $z=0$, in the plane of Fig. \ref{fig:rho_SPIRAL6} in four zones, two for each bubbles. The two pale-blue regions to the right of the center that are surrounded by the denser green regions and with a faster outflow velocity than the surroundings show the cross section of the bubbles' segments  (one above and one below the equatorial plane $z=0$) that the jets inflated at earlier times. The green zones to the left of the center that are surrounded by yellow (higher density) regions and which have very high velocities are segments of the bubbles that the NS inflated recently. At the time of this figure ($t=3.2 \yr$) the NS is at $(x,y,z)_{\rm NS}=(-1.05 \times 10^{13} \cm, 1.8 \times 10^{13} \cm, 0)$.  

The outflow pattern while the NS is still in the outer regions of the envelope is similar to our earlier studies where the NS had a fixed orbital separation in the outer envelope (e.g., \citealt{Hilleletal2022FB, Schreieretal2023}).
The morphology of the ejected envelope at these early times  is qualitatively similar to that in the GEE (e.g., \citealt{Shiberetal2017, ShiberSoker2018}). We here do not study the outflow morphology but rather examine the vortices that the jets induce. 

\begin{figure}  
\centering
\includegraphics[width=0.42\textwidth]{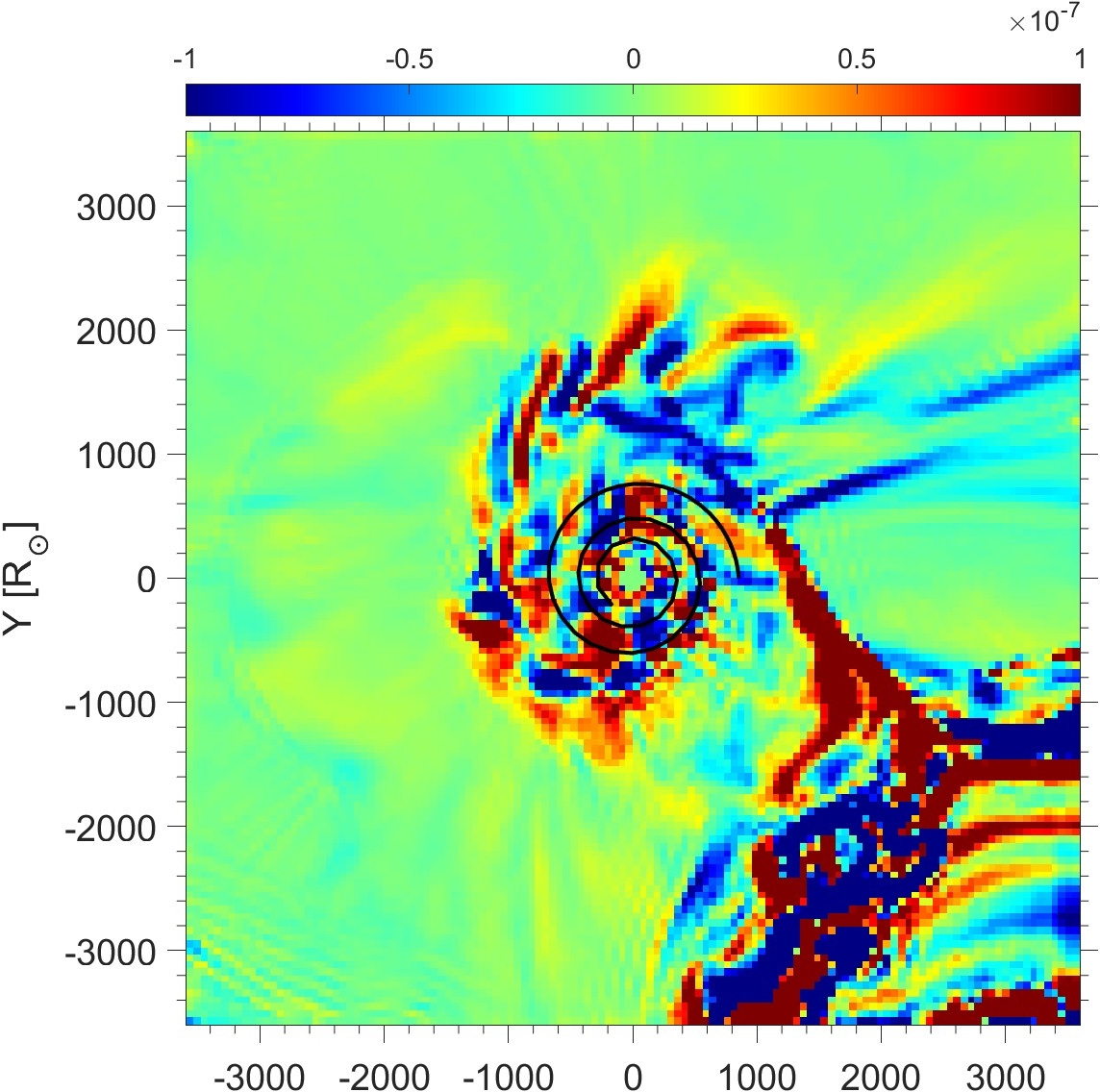} 
\includegraphics[width=0.42\textwidth]{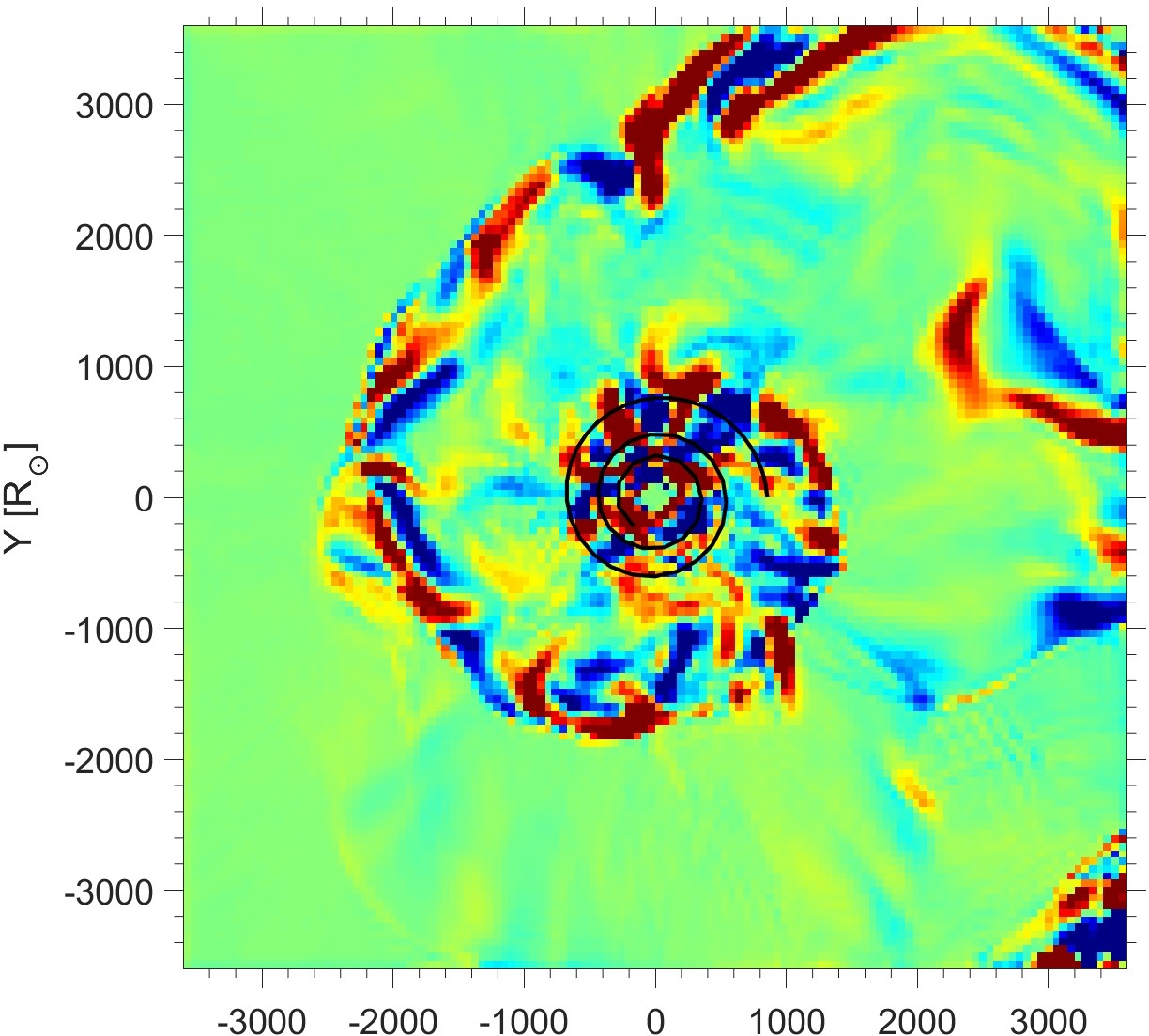}\\
\includegraphics[width=0.42\textwidth]{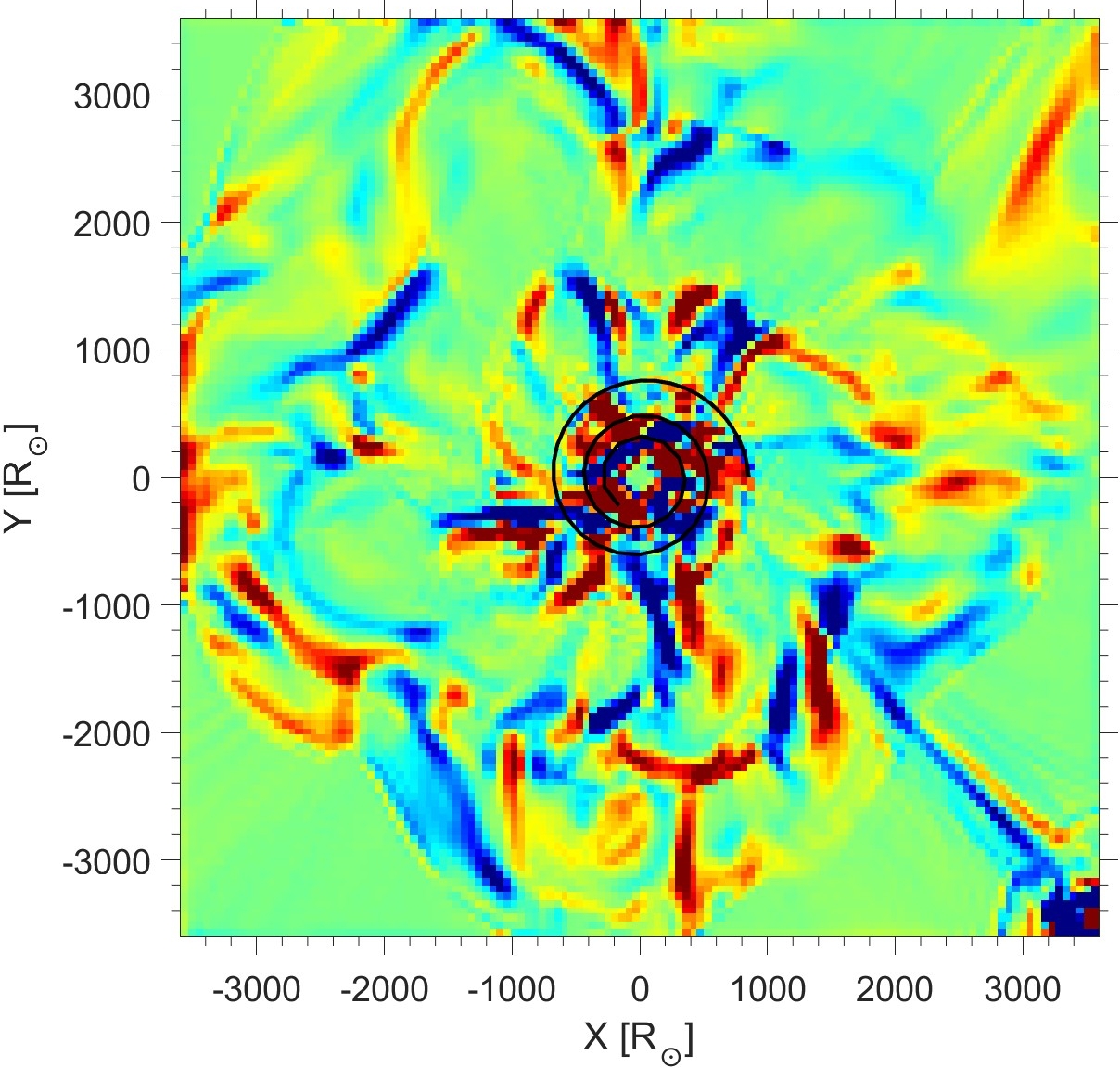}
\caption{
The quantity $(\overrightarrow \nabla \times \overrightarrow v)_z$ in the $z= 6 \times 10^{12} \cm=86R_\odot$ plane for the simulation with jets. Values according to the color bar in units of $\s^{-1}$. From top to bottom are plots at $t=2 \yr$ just as the NS reaches an orbital separation of $a= 500 R_\odot$,
$t=3 \yr$ when the NS reaches $a_{\rm SR} = 300 R_\odot$,  and $t=4 \yr$ as the NS continues to orbit at $a_{\rm SR}=300 R_\odot$. 
The black spiral in the centre indicates the trajectory of the spiralling-in NS.
}
\label{fig:Vortices3Times}
\end{figure}

In addition to the large scale flow that inflate the bubbles and sets an outflow, we notice the vortices that the jets induce inside the envelope. To better reveal the vortices we present in Fig. \ref{fig:Vortices3Times} the $z$ component of the curl of the velocity, $(\overrightarrow \nabla \times \overrightarrow v)_z$, in the $z= 6 \times 10^{12} \cm$ plane. This figure shows that the jets shed  pairs of vortices with opposite signs as the NS spirals-in inside the RSG envelope. The typical width of each vortex with the typical value of $\vert (\overrightarrow \nabla \times \overrightarrow v)_z \vert \simeq 10^{-7} \s^{-1}$ is $d_{\rm v} \simeq 10^{13} \cm$. The corresponding typical velocity of gas in the vortices is $v_{\rm v} \simeq 10 \km \s^{-1}$. 
Some smaller regions within the above regions have higher values of $\vert \overrightarrow \nabla \times \overrightarrow v)_z \simeq 3 \times 10^{-7} \s^{-1}$ and in these smaller regions $v_{\rm v} \simeq 20-30 \km \s^{-1}$

\begin{figure*} 
\centering
\includegraphics[width=0.45\textwidth]{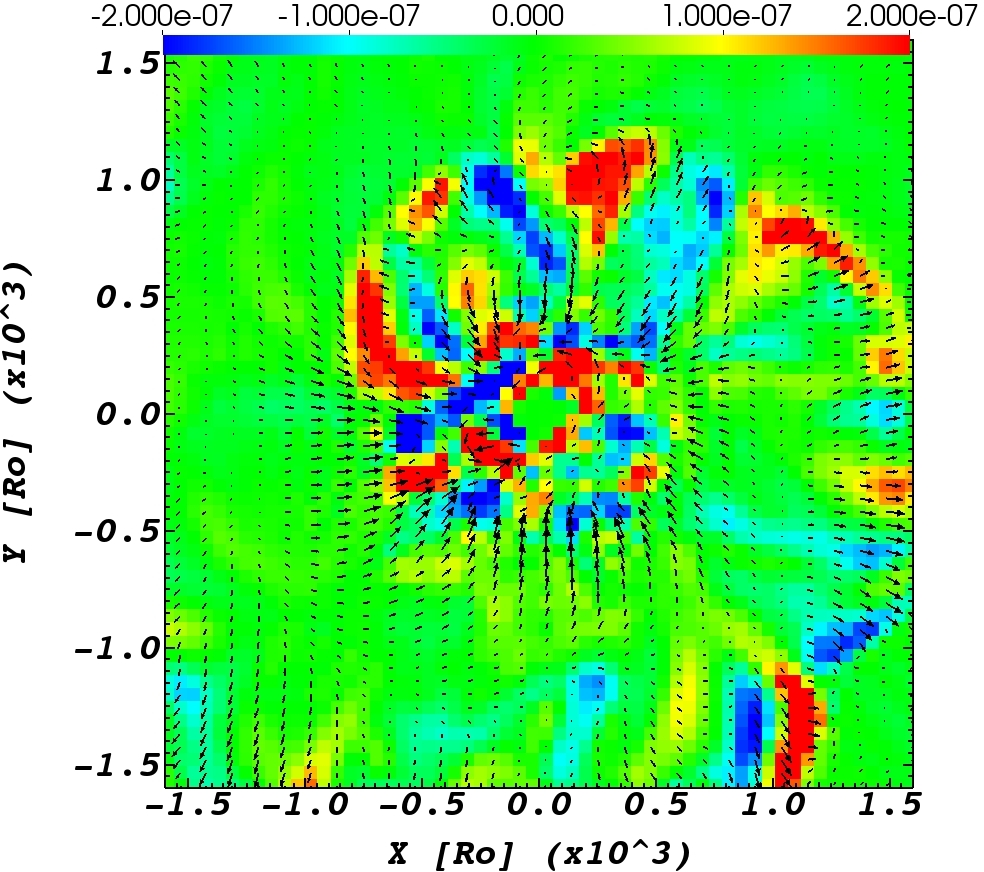}
\includegraphics[width=0.45\textwidth]{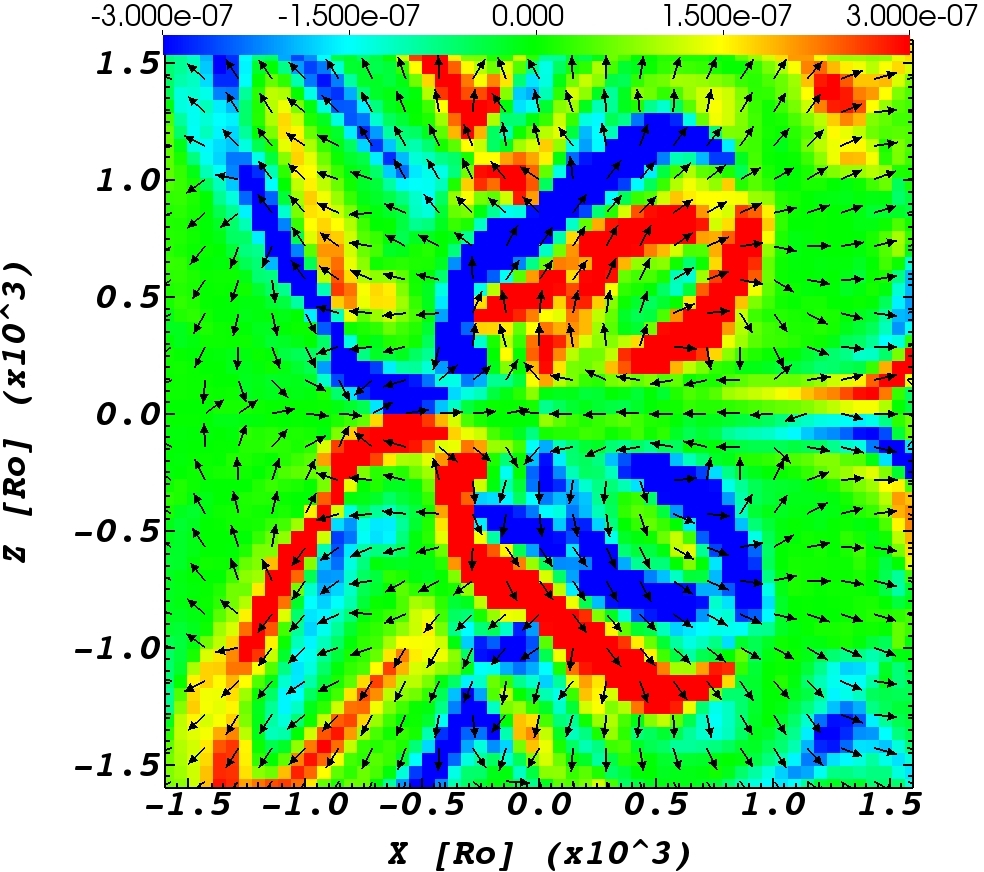} \\
\includegraphics[width=0.45\textwidth]{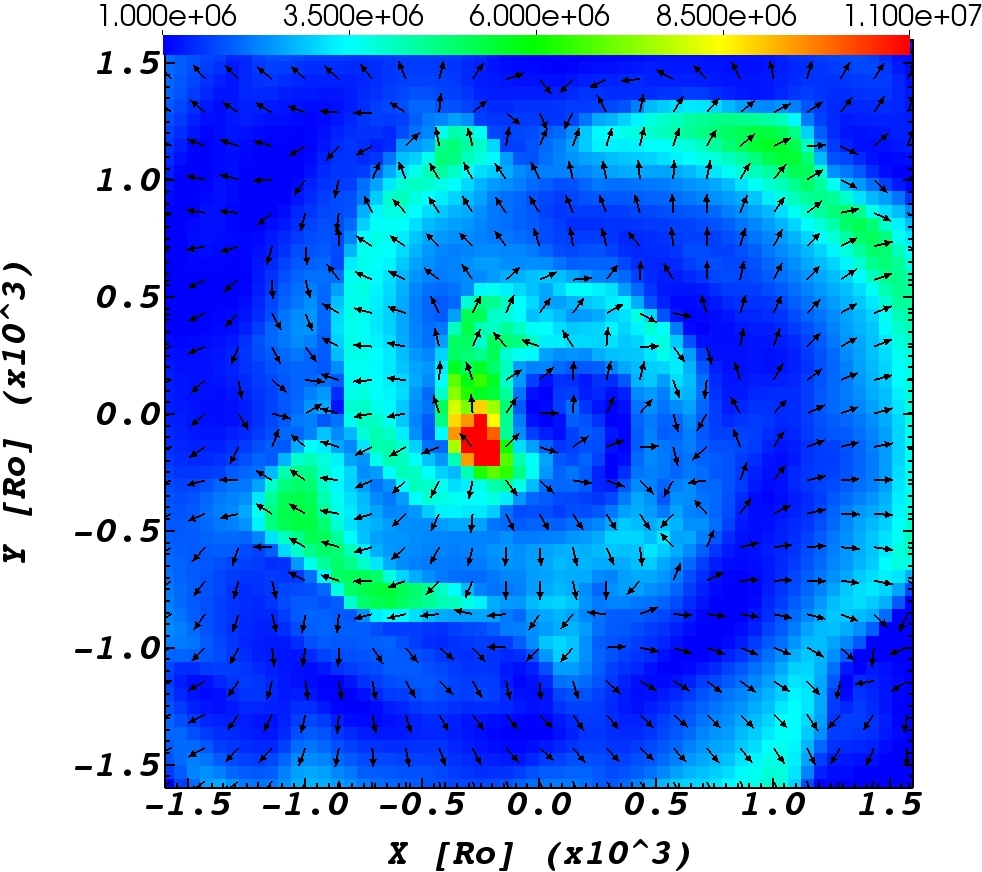}
\includegraphics[width=0.45\textwidth]{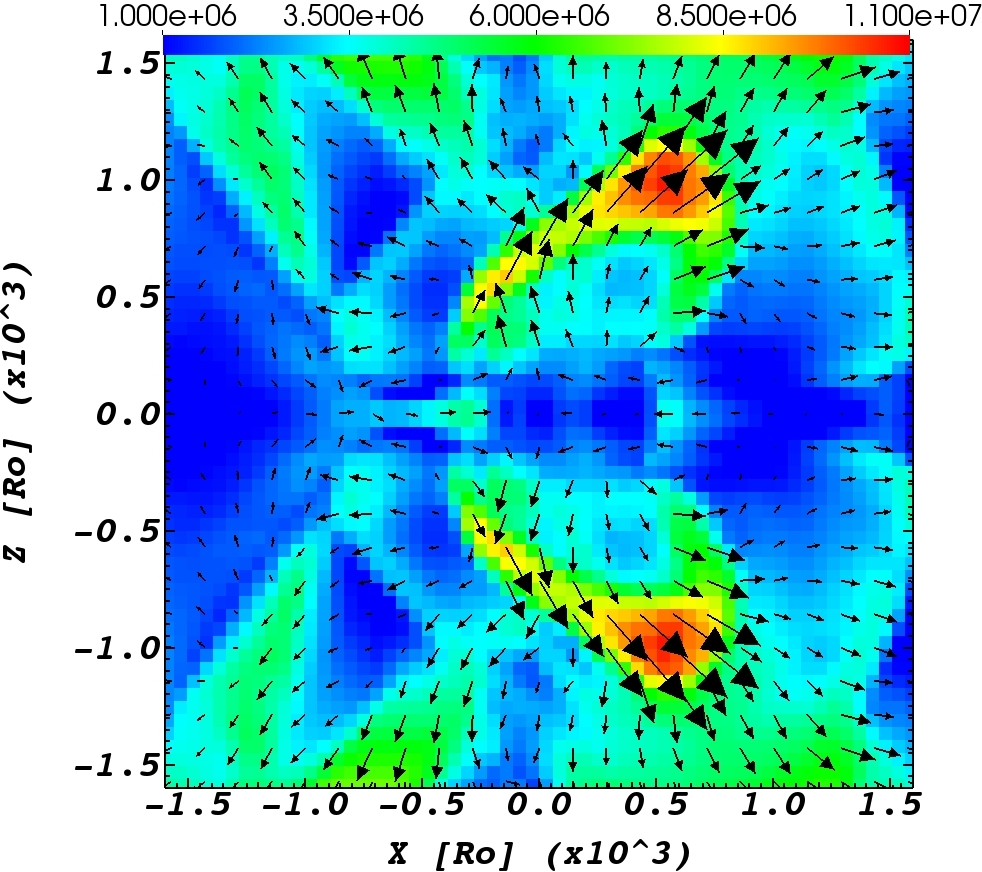}
\caption{ Upper panels: 
Velocity arrows on top of the curl of the velocity map $(\overrightarrow \nabla \times \overrightarrow v)_z$ in the $z=6 \times 10^{12} \cm = 86 R_\odot$ plane
(left) and of $(\overrightarrow \nabla \times \overrightarrow v)_y$ in the plane $y=1.8\times 10^{13} \cm = 259 R_\odot$ (right), at $t=3.2 \yr$  (units of color bar are $\s^{-1}$). 
Lower panels: Velocity arrows on top of velocity magnitude according to the color bar (units of  $\cm \s^{-1}$) and in the same planes and the same time as in the upper panels. 
Maximum velocity in the plane on the left is $ 40 \km \s^{-1}$ and in the plane on the right $105 \km \s^{-1}$. The high velocity flow on both sides of the equatorial plane in the right panels are due to the two jets launched by the NS, which at this time is at 
$(x,y,z)_{\rm NS}=(151 R_\odot, 259 R_\odot, 0)$.  
}
\label{fig:curl_vel_vec_XY_XZ}
\end{figure*}

To further explore the properties of the vortices we present in Fig. \ref{fig:curl_vel_vec_XY_XZ} the vortices in the inner region of our numerical grid. In the upper-left panel we present $(\overrightarrow \nabla \times \overrightarrow v)_z$ in the $z=6\times10^{12} \cm$ plane and in the upper-right panel we present $(\overrightarrow \nabla \times \overrightarrow v)_y$ in the  plane $y=1.8 \times 10^{13} \cm$. The vortices have larger values of $(\overrightarrow \nabla \times \overrightarrow v)_y$ than of $(\overrightarrow \nabla \times \overrightarrow v)_z$. This is because the jets are injected along the $z$ direction. In the lower two panels we present the velocity directions by arrows on top of the velocity magnitude coded by the colour. The time and planes are as in the upper panels.  The outflow velocities reach values of up to $v_{\rm out} \simeq 100 \km \s^{-1}$. Overall, the outflow velocities are higher than the circularisation velocity in the vortices.
Nonetheless, there are small regions where there is an inflow. As we discuss in section \ref{sec:Summary}, these inflowing regions might lead to fall back material at late phases of the evolution.

\section{Deposition of angular momentum} 
\label{sec:AngularMomentum}
  
We calculate the angular momentum that the jets deposit to the envelope as in \cite{Schreieretal2023}. We take the envelope to be the gas between the inert core at $r_{\rm inert} = 0.2 R_{\rm RSG} = 1.23 \times 10^{13} \cm = 177 R_\odot$ and an outer radius of $r = 1.25 \times 10^{14} \cm = 1796 R_\odot$ which is about twice the initial radius of the RSG star. Namely, in calculating the envelope angular momentum we sum the quantity $m_i \vec{r_i} \times \vec{v_i}$ over all cells $i$ with $r_{\rm inert} < r_i < 1.25 \times 10^{14} \cm$, where $m_i$, $\vec{r_i}$ and $\vec{v_i}$ are the mass, the location with respect to the centre of the RSG, and the velocity of cell $i$.
We recall that we do not include the orbital angular momentum, nor the gravity of the NS. Therefore, only jets influence the envelope as our goal is to explore the role of jets. 

\begin{figure} 
\centering
\includegraphics[width=0.45\textwidth]{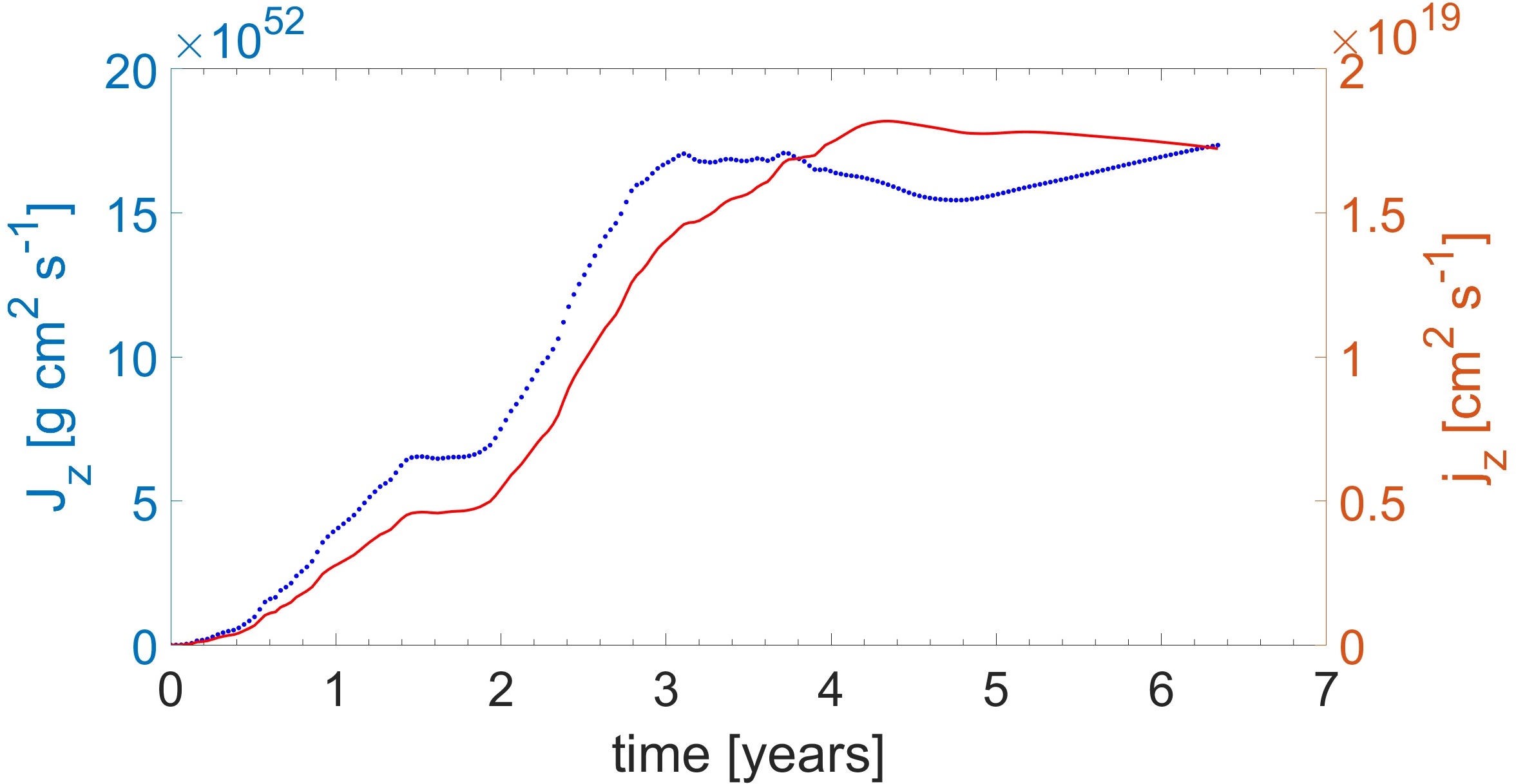}
\caption{
The angular momentum that the jets deposit to the envelope as function of time (dotted-blue line, scale of the left) and the specific angular momentum in the envelope (solid-red line, scale on the right) as function of time.  }
\label{fig:J_z}
\end{figure}

In Fig. \ref{fig:J_z} we present the angular momentum that the jet deposit to the envelope $J_z$, and the specific angular momentum in the envelope, $j_z=J_z/M_{\rm env}$, as function of time. The increase in angular momentum occurs while the NS, which is the source of the jets, spirals-in. At $t>3 \yr$ the NS continues to orbit the core of the RSG star at a constant radius of $a_{\rm SR}=300 R_\odot$.  The increase in $J_z$ (and in $j_z$) is not monotonic. There is a time period when the angular momentum stays almost constant for about half a year, $t\simeq 1.4-1.9 \yr$, before the NS reaches its final orbit. This time corresponds more or less to the time when the NS completes an orbital angle of $\simeq 360^\circ$.   

In our previous study \citep{Schreieretal2023} we simulated jets with a constant orbital radius of $a=700 R_\odot=4.9 \times 10^{13} \cm$ and followed the angular momentum that the jets deposit to the envelope. We found that when the jets are perpendicular to the orbital plane the jets deposit substantial amount of angular momentum to the envelope only in the first two orbits. Based on that we suggested that jets deposit angular momentum to the envelope only during the plunge-in phase, when the NS rapidly spirals-in, but only small amounts during the self-regulated phase when the spiralling-in is very slow or does not take place at all. Our results here where we let the NS to plunge-in in three years confirm our expectation. 

The deposition of angular momentum, therefore, results from the asymmetrical flow structure that occurs whenever the NS enters a new envelope shell that was not perturbed directly before. We can quantify the departure from symmetry. The average angular momentum deposition rate during the spiralling-in phase is $\dot J_{\rm jets} \simeq 1.8 \times 10^{53} \g \cm^2 \s^{-1}/3 \yr \simeq 2 \times 10^{45} \g \cm^2 \s^{-2}$. Within this three-year plunge-in time period the jets deposit energy of $E _{\rm jets,t} \simeq 2 \times 10^{49} \erg$. The ratio of these quantities is $\eta_{jz} \equiv \dot J_{\rm jets}/ E _{\rm jets,t} \simeq 10^{-4}$. This is the degree of asymmetry of the flow structure that is powered by the jets and that leads to envelope-spin-up along the orbital angular momentum axis. The jets induce much larger departures from spherical symmetry in the envelope, but $\eta_{jz}$ refers only to angular momentum deposition.       
  
The total angular momentum that the jets deposit to the envelope at the end of the plunge-in phase here ($t=3 \yr$) is $J_{\rm jets} \simeq 1.8 \times 10^{53} \g \cm^2 \s^{-1}$. In addition, the spiralling-in NS deposits orbital angular momentum to the envelope, a process we do not simulate here. In spiralling-in from $a_0=850 R_\odot$ to the final orbit during the self-regulated phase at $a_{\rm SR} = 300 R_\odot$ the decrease in the orbital angular momentum is $J_{\rm orb} = 4.5 \times 10^{53} \g \cm^2 \s^{-1}$. Considering that the jets in reality might be more powerful than what we simulate (which we cannot numerically simulate with our resources due to too small time steps), we conclude that the contribution of jets to spinning-up the envelope is substantial, and might event dominate. 

\section{Summary} 
\label{sec:Summary}

This study continues our exploration of the roles that jets that a NS launches play in CEE. Our numerical resources do not allow for high-resolution simulations with jets. In section \ref{sec:StellarModel} we examined the behaviour of our 3D RSG stellar model without any jets, which does allow for high resolution. We found that the stellar model, in both the regular and the high-resolution simulations, performs two non-linear oscillations before it relaxes in a time of about 3 years (Figs. \ref{fig:DensityMaps} and \ref{fig:ShellsRadii}). Additionally, turbulence is developed in the envelope (Figs. \ref{fig:TracerMaps}, \ref{fig:VorticiesCuts_LR} and \ref{fig:VorticiesCuts_HR}).  

Both the non-linear oscillations and the turbulence are processes that take place in giant stars (RGB, AGB, RSG). Because we have no source of nuclear burning in the centre, the oscillations decay. We concluded in section \ref{sec:StellarModel} that in 3D hydrodynamical simulations of the CEE with giant stars there is no need to stabilise the giant star to a high degree when transferring a 1D model to a 3D grid. One can start the binary simulations immediately, even if the giant model is not completely static. Just let it oscillate and develop turbulence, as such giants should actually have. 

We then presented the results of a simulation with regular resolution where the NS, that serves here only as the source of the jets as we do not include its gravity (section \ref{sec:Numerical}), to spiral-in with a predetermined orbit (spiral line in the three panels of Fig. \ref{fig:Vortices3Times}). The exclusion of the NS gravity as in our earlier studies (e.g., \citealt{Hilleletal2022FB, Schreieretal2023}) allows us to identify the role of jets and to perform the simulations on our computer. 

The new ingredient of this study is the spiralling-in of the NS.
The jets inflate two expanding spiral-shaped low-density bubbles, one above and one below the equatorial plane. The two cross sections of each bubble with the $y=0$ plane are seen in Fig. \ref{fig:rho_SPIRAL6} as two low-density high-velocity regions above (one bubble) and two below (the other bubble) the equatorial plane $z=0$.  

We also find that the NS sheds pairs of opposite-sign vortices as it spirals-in, best seen as blue-red  pairs in Figs. \ref{fig:Vortices3Times} and \ref{fig:curl_vel_vec_XY_XZ}. The pairs of vortices form an expanding large-scale spiral pattern.  Overall, the jets substantially increase the turbulence in the common envelope, both as random motion on small scales and as a global pattern that substantially deviates from a spherical structure. We emphasise that the spiral structure seen in and near the equatorial plane as pairs of vortices (Fig. \ref{fig:Vortices3Times} and upper-left panel of Fig. \ref{fig:curl_vel_vec_XY_XZ}) and in the density (lower-left panel of Fig. \ref{fig:curl_vel_vec_XY_XZ}) are formed solely by jets, as we do not include the NS gravity. 

Convection in the envelope of red giants can efficiently transport angular momentum (e.g., \citealt{GagnierPejcha2023}) and energy (e.g., \citealt{Gricheneretal2018, WilsonNordhaus2019, WilsonNordhaus2020, WilsonNordhaus2022}). We showed here that the jets can substantially increase the convection (turbulence) strength. This makes energy transport more efficient. Namely, a fraction of the energy that the jets deposit to the envelope is carried away and radiated. The transient event, termed CEJSN, is very bright, lasts for months to a few years, and might mimic a very energetic core collapse supernova.

The jet-induced non-spherical morphology of the ejected envelope influences the late light curve. This occurs if at later times inner envelope gas is ejected at higher velocities and collides with early ejecta. This collision converts kinetic energy to thermal energy and then radiation. The non-spherical structure leads to bumps in the light curve and to polarised emission. 

We also confirm our suggestion from  \cite{Schreieretal2023} that the jets deposit a non-negligible amount of angular momentum to the envelope during the plunge-in phase, when the spiralling-in is on a dynamical timescale, but not much during the self-regulated phase when the spiralling-in is very slow. The direction of the angular momentum that the jets deposit is the same as that of the orbital angular momentum. The deposition of positive angular momentum results from the fact that the jets eject envelope mass with negative angular momentum. 

We do not resolve the accretion process and the propagation of the jets through the accretion flow. These processes are simulated by \cite{MorenoMendezetal2017}, \cite{LopezCamaraetal2019}, and \cite{LopezCamaraetal2020MN}. They do not follow the spiralling-in orbit. The main common feature is that both types of simulations find the jets to inflate bubbles. \cite{MorenoMendezetal2017} simulate the jets in a box (not in a spiralling-in orbit), and find that the jets that expand into the envelope inflate bubbles, one at each side of the equatorial plane (also \citealt{LopezCamaraetal2019}). We find each bubble to possess a spiral structure with its cross section increasing with distance from the center as the gas in the bubble expands out. \cite{MorenoMendezetal2017} determine the conditions for the jets to propagate into the common envelope. An important result of \cite{MorenoMendezetal2017} is that in many cases the jets are expected to be intermittent. We expect that in that case the angular momentum deposition into the envelope might be larger than what we find in our study because intermittent activity increases the asymmetry of the flow structure around the angular momentum axis. 

This study supports the general claim that jets that NSs (and BHs) launch during CEE, namely during a CEJSN event, cannot be neglected. Jets power the CEJSN event, influence the morphology of the ejected envelope, induce vortices that strengthen the convection, and deposit angular momentum to the common envelope. 

\section*{Acknowledgments}
We thank Diego L{\'o}pez-C{\'a}mara and an anonymous referee for helpful comments.
This research was supported by the Amnon Pazy Research Foundation.

\section*{Data availability}
The data underlying this article will be shared on reasonable request to the corresponding author.  


\label{lastpage}


\begin{thebibliography}{}\addcontentsline{toc}{section}{References}

\bibitem[Armitage \& Livio(2000)]{ArmitageLivio2000} Armitage, P.~J., \& Livio, M.\ 2000, \apj, 532, 540

\bibitem[\protect\citeauthoryear{Blackman \& Lucchini}{2014}]{BlackmanLucchini2014} Blackman E.~G., Lucchini S., 2014, MNRAS, 440, L16.

\bibitem[\protect\citeauthoryear{Chamandy et al.}{2019}]{Chamandyetal2019} Chamandy L., Blackman E.~G., Frank A., Carroll-Nellenback J., Zou Y., Tu Y., 2019, MNRAS, 490, 3727.

\bibitem[\protect\citeauthoryear{Chamandy et al.}{2018}]{Chamandyetal2018} Chamandy L., Frank A., Blackman E.~G., Carroll-Nellenback J., Liu B., Tu Y., Nordhaus J., et al., 2018, MNRAS, 480, 1898.

\bibitem[\protect\citeauthoryear{Chamandy et al.}{2023}]{Chamandyetal2023} Chamandy L., Carroll-Nellenback J., Blackman E.~G., Frank A., Tu Y., Liu B., Zou Y., et al., 2023,  arXiv:2304.14840.

\bibitem[Chevalier(2012)]{Chevalier2012} Chevalier, R.~A.\ 2012, \apjl, 752, L2

\bibitem[\protect\citeauthoryear{Fields \& Couch}{2020}]{FieldsCouch2020} Fields C.~E., Couch S.~M., 2020, ApJ, 901, 33.

\bibitem[\protect\citeauthoryear{Fields \& Couch}{2021}]{FieldsCouch2021} Fields C.~E., Couch S.~M., 2021, ApJ, 921, 28.

\bibitem[Fryxell et al.(2000)] {Fryxelletal2000} Fryxell, B., et al. 2000, \apjs, 131, 273

\bibitem[\protect\citeauthoryear{Gagnier \& Pejcha}{2023}]{GagnierPejcha2023} Gagnier D., Pejcha O., 2023, A\&A, 674, A121. 

\bibitem[\protect\citeauthoryear{Gilkis \& Soker}{2016}]{GilkisSoker2016} Gilkis A., Soker N., 2016, ApJ, 827, 40.

\bibitem[\protect\citeauthoryear{Glanz \& Perets}{2021a}]{GlanzPerets2021a} Glanz H., Perets H.~B., 2021a, MNRAS, 500, 1921.

\bibitem[\protect\citeauthoryear{Glanz \& Perets}{2021b}]{GlanzPerets2021b} Glanz H., Perets H.~B., 2021b, MNRAS, 507, 2659.

\bibitem[\protect\citeauthoryear{Gonz{\'a}lez-Bol{\'\i}var et al.}{2022}]{GonzalezBolivar2022} Gonz{\'a}lez-Bol{\'\i}var M., De Marco O., Lau M.~Y.~M., Hirai R., Price D.~J., 2022, MNRAS, 517, 3181.

\bibitem[\protect\citeauthoryear{Grichener, Cohen, \& Soker}{2021}]{Gricheneretal2021} Grichener A., Cohen C., Soker N., 2021, ApJ, 922, 61.

\bibitem[\protect\citeauthoryear{Grichener, Sabach, \& Soker}{2018}]{Gricheneretal2018} Grichener A., Sabach E., Soker N., 2018, MNRAS, 478, 1818.

\bibitem[\protect\citeauthoryear{Hillel, Schreier, \& Soker}{2022}]{Hilleletal2022FB} Hillel S., Schreier R., Soker N., 2022, MNRAS, 514, 3212.

\bibitem[Iaconi et al.(2017b)]{Iaconietal2017b} Iaconi, R., Reichardt, T., Staff, J., De Marco, O., Passy, J.-C., Price, D., Wurster, J., \& Herwig, F.\ 2017b, \mnras, 464, 4028

\bibitem[Kuruwita et al.(2016)]{Kuruwitaetal2016} Kuruwita, R.~L., Staff, J., \& De Marco, O.\ 2016, \mnras, 461, 486

\bibitem[\protect\citeauthoryear{Lau et al.}{2022a}]{Lauetal2022a} Lau M.~Y.~M., Hirai R., Gonz{\'a}lez-Bol{\'\i}var M., Price D.~J., De Marco O., Mandel I., 2022a, MNRAS, 512, 5462.

\bibitem[\protect\citeauthoryear{Lau et al.}{2022b}]{Lauetal2022b} Lau M.~Y.~M., Hirai R., Price D.~J., Mandel I., 2022b, MNRAS, 516, 4669.

\bibitem[\protect\citeauthoryear{Law-Smith et al.}{2020}]{LawSmithetal2020} Law-Smith J.~A.~P., Everson R.~W., Ramirez-Ruiz E., de Mink S.~E., van Son L.~A.~C., G{\"o}tberg Y., Zellmann S., et al., 2020, arXiv:2011.06630

\bibitem[L{\'o}pez-C{\'a}mara et al.(2019)]{LopezCamaraetal2019} L{\'o}pez-C{\'a}mara, D., De Colle, F., \& Moreno M{\'e}ndez, E.\ 2019, \mnras, 482, 3646

\bibitem[\protect\citeauthoryear{Lopez-Camara et al.}{2022}]{LopezCamaraetal2022} L{\'o}pez-C{\'a}mara D., De Colle F., Moreno M{\'e}ndez E., Shiber S., Iaconi R., 2022, MNRAS, 513, 3634.

\bibitem[L{\'o}pez-C{\'a}mara et al.(2020)]{LopezCamaraetal2020MN} L{\'o}pez-C{\'a}mara, D., Moreno M{\'e}ndez, E., \& De Colle, F.\ 2020, \mnras, 497, 2057

\bibitem[\protect\citeauthoryear{Moreno et al.}{2022}]{Morenoetal2022} Moreno M.~M., Schneider F.~R.~N., R{\"o}pke F.~K., Ohlmann S.~T., Pakmor R., Podsiadlowski P., Sand C., 2022, A\&A, 667, A72.

\bibitem[Moreno M{\'e}ndez et al.(2017)]{MorenoMendezetal2017} Moreno M{\'e}ndez, E., L{\'o}pez-C{\'a}mara, D., \& De Colle, F.\ 2017, \mnras, 470, 2929

\bibitem[Nandez et al.(2014)]{Nandezetal2014} Nandez, J.~L.~A., Ivanova, N., \& Lombardi, J.~C., Jr.\ 2014, \apj, 786, 39

\bibitem[\protect\citeauthoryear{Ohlmann et al.}{2016}]{Ohlmannetal2016a} Ohlmann S.~T., R{\"o}pke F.~K., Pakmor R., Springel V., 2016, ApJL, 816, L9.

\bibitem[\protect\citeauthoryear{Ondratschek et al.}{2022}]{Ondratscheketal2022} Ondratschek P.~A., R{\"o}pke F.~K., Schneider F.~R.~N., Fendt C., Sand C., Ohlmann S.~T., Pakmor R., et al., 2022, A\&A, 660, L8.

\bibitem[Passy et al.(2012)]{Passyetal2012} Passy, J.-C., De Marco, O., Fryer, C.~L., et al.\ 2012, \apj, 744, 52

\bibitem[Paxton et al.(2011)]{Paxtonetal2011} Paxton, B., Bildsten, L., Dotter, A., et al.\ 2011, \apjs, 192, 3

\bibitem[Paxton et al.(2013)] {Paxtonetal2013} Paxton, B., Cantiello, M., Arras, P., et al. 2013, \apjs, 208, 4

\bibitem[Paxton et al.(2015)]{Paxtonetal2015} Paxton, B., Marchant, P., Schwab, J., et al.\ 2015, \apjs, 220, 15

\bibitem[Paxton et al.(2018)]{Paxtonetal2018} Paxton, B., Schwab, J., Bauer, E.~B., et al.\ 2018, \apjs, 234, 34

\bibitem[Paxton et al.(2019)]{Paxtonetal2019} Paxton, B., Smolec, R., Schwab, J., et al.\ 2019, \apjs, 243, 10, 

\bibitem[Ricker \& Taam(2012)]{RickerTaam2012} Ricker, P.~M., \& Taam, R.~E.\ 2012, \apj, 746, 74

\bibitem[\protect\citeauthoryear{Roepke \& De Marco}{2023}]{RoepkeDeMarco2023} Roepke F.~K., De Marco O., 2023, Living Rev Comput Astrophys, 9, 2

\bibitem[Schreier et al.(2019)]{Schreieretal2019inclined} Schreier, R., Hillel, S., \& Soker, N.\ 2019, \mnras, 490, 4748.

\bibitem[\protect\citeauthoryear{Schreier, Hillel, \& Soker}{2023}]{Schreieretal2023} Schreier R., Hillel S., Soker N., 2023, MNRAS, 520, 4182.

\bibitem[\protect\citeauthoryear{Shiber et al.}{2019}]{Shiberetal2019} Shiber S., Iaconi R., De Marco O., Soker N., 2019, MNRAS, 488, 5615.

\bibitem[\protect\citeauthoryear{Shiber, Kashi, \& Soker}{2017}]{Shiberetal2017} Shiber S., Kashi A., Soker N., 2017, MNRAS, 465, L54.

\bibitem[\protect\citeauthoryear{Shiber, Schreier, \& Soker}{2016}]{Shiberetal2016} Shiber S., Schreier R., Soker N., 2016, RAA, 16, 117.

\bibitem[Shiber \& Soker(2018)]{ShiberSoker2018} Shiber, S. \& Soker, N.\ 2018, \mnras, 477, 2584

\bibitem[\protect\citeauthoryear{Soker}{1992}]{Soker1992} Soker N., 1992, ApJ, 389, 628.

\bibitem[Soker(2016b)]{Soker2016Rev} Soker N., 2016, NewAR, 75, 1.

\bibitem[\protect\citeauthoryear{Soker}{2022}]{Soker2022Rev} Soker N., 2022, RAA, 22, 122003.

\bibitem[\protect\citeauthoryear{Soker}{2023}]{Soker2023MS} Soker N., 2023, arXiv:2303.05880.

\bibitem[Staff et al.(2016a)]{Staffetal2016MN} Staff, J.~E., De Marco, O., Macdonald, D., Galaviz, P., Passy, J.C., Iaconi, R., \& Mac Low, M.-M\ 2016a, \mnras, 455, 3511

\bibitem[\protect\citeauthoryear{Trabucchi et al.}{2021}]{Trabucchietal2021} Trabucchi M., Wood P.~R., Mowlavi N., Pastorelli G., Marigo P., Girardi L., Lebzelter T., 2021, MNRAS, 500, 1575.

\bibitem[\protect\citeauthoryear{Wilson \& Nordhaus}{2019}]{WilsonNordhaus2019} Wilson E.~C., Nordhaus J., 2019, MNRAS, 485, 4492.

\bibitem[\protect\citeauthoryear{Wilson \& Nordhaus}{2020}]{WilsonNordhaus2020} Wilson E.~C., Nordhaus J., 2020, MNRAS, 497, 1895.

\bibitem[\protect\citeauthoryear{Wilson \& Nordhaus}{2022}]{WilsonNordhaus2022} Wilson E.~C., Nordhaus J., 2022, MNRAS, 516, 2189.

\bibitem[\protect\citeauthoryear{Zou et al.}{2022}]{Zouetal2022} Zou Y., Chamandy L., Carroll-Nellenback J., Blackman E.~G., Frank A., 2022, MNRAS, 514, 3041.

\bibitem[\protect\citeauthoryear{Zou et al.}{2020}]{Zouetal2020} Zou Y., Frank A., Chen Z., Reichardt T., De Marco O., Blackman E.~G., Nordhaus J., et al., 2020, MNRAS, 497, 2855.

\end{thebibliography}
\end{document}